\documentclass[traditabstract,longauth]{aa} 
\usepackage{graphicx}
\usepackage{amsmath}
\usepackage{txfonts}
\usepackage{natbib}                
\bibpunct{(}{)}{;}{a}{}{,}         
\usepackage{verbatim}
\usepackage{fixltx2e} 
\usepackage{multirow}
\newcommand{\pks}{PKS\,0447$-$439 }

\newcommand{\fermi}{\textit{Fermi}-LAT}

\begin{document}

%
   \title{Discovery of TeV $\gamma$-ray emission from PKS\,0447-439 and derivation of an upper limit on its redshift}

   \titlerunning{Discovery of VHE $\gamma$-ray emission from PKS~0447-439}

\author{H.E.S.S. Collaboration
\and A.~Abramowski \inst{1}
\and F.~Acero \inst{2}
\and A.G.~Akhperjanian \inst{6,5}
\and G.~Anton \inst{7}
\and S.~Balenderan \inst{8}
\and A.~Balzer \inst{9,10}
\and A.~Barnacka \inst{11,12}
\and Y.~Becherini \inst{13,14}
\and J.~Becker Tjus \inst{15}
% ...
\newpage
}

\offprints{\\Andreas.Zech@obspm.fr, Bagmeet.Behera@desy.de}

\institute{
Universit\"at Hamburg, Institut f\"ur Experimentalphysik, Luruper Chaussee 149, D 22761 Hamburg, Germany \and
Laboratoire Univers et Particules de Montpellier, Universit\'e Montpellier 2, CNRS/IN2P3,  CC 72, Place Eug\`ene Bataillon, F-34095 Montpellier Cedex 5, France \and
Max-Planck-Institut f\"ur Kernphysik, P.O. Box 103980, D 69029 Heidelberg, Germany \and
Dublin Institute for Advanced Studies, 31 Fitzwilliam Place, Dublin 2, Ireland \and
National Academy of Sciences of the Republic of Armenia, Yerevan  \and
Yerevan Physics Institute, 2 Alikhanian Brothers St., 375036 Yerevan, Armenia \and
Universit\"at Erlangen-N\"urnberg, Physikalisches Institut, Erwin-Rommel-Str. 1, D 91058 Erlangen, Germany \and
University of Durham, Department of Physics, South Road, Durham DH1 3LE, U.K. \and
DESY, D-15735 Zeuthen, Germany \and
Institut f\"ur Physik und Astronomie, Universit\"at Potsdam,  Karl-Liebknecht-Strasse 24/25, D 14476 Potsdam, Germany \and
Nicolaus Copernicus Astronomical Center, ul. Bartycka 18, 00-716 Warsaw, Poland \and
CEA Saclay, DSM/Irfu, F-91191 Gif-Sur-Yvette Cedex, France \and
APC, AstroParticule et Cosmologie, Universit\'{e} Paris Diderot, CNRS/IN2P3, CEA/Irfu, Observatoire de Paris, Sorbonne Paris Cit\'{e}, 10, rue Alice Domon et L\'{e}onie Duquet, 75205 Paris Cedex 13, France,  \and
Laboratoire Leprince-Ringuet, Ecole Polytechnique, CNRS/IN2P3, F-91128 Palaiseau, France \and
Institut f\"ur Theoretische Physik, Lehrstuhl IV: Weltraum und Astrophysik, Ruhr-Universit\"at Bochum, D 44780 Bochum, Germany
% ...
}

\date{Received ...; accepted ...}

  \abstract
    {Very high-energy $\gamma$-ray emission from \pks was detected with the H.E.S.S. Cherenkov telescope array in December 2009. This blazar is one of the brightest extragalactic objects in the \emph{Fermi} Bright Source List and has a hard spectrum in the MeV to GeV range. In the TeV range, a photon index of 3.89 $\pm$ 0.37 (stat) $\pm$ 0.22 (sys) and a flux normalisation at 1\,TeV, $\Phi_{1\mathrm{TeV}}=(3.5\pm1.1 (stat)\pm0.9 (sys) )\times10^{-13}\mathrm{cm}^{-2}\mathrm{s}^{-1}\mathrm{TeV}^{-1}$ were found. The detection with H.E.S.S. triggered observations in the X-ray band with the \emph{Swift} and \emph{RXTE} telescopes. Simultaneous UV and optical data from \emph{Swift} UVOT and data from the optical telescopes ATOM and ROTSE are also available. 
   The spectrum and light curve measured with H.E.S.S. are presented and compared to the multi-wavelength data at lower energies. A rapid flare is seen in the \emph{Swift} XRT and \emph{RXTE} data, 
   together with a flux variation in the UV band, at a time scale of the order of one day. 
  A firm upper limit of z $<$\,0.59 on the redshift of \pks is derived from the combined \emph{Fermi}-LAT and H.E.S.S. data, given the assumptions that there is no upturn in the intrinsic spectrum above the \emph{Fermi}-LAT energy range and that absorption on the Extragalactic Background Light (EBL) is not weaker than the lower limit provided by current models. The spectral energy distribution is well described by a simple one-zone Synchrotron Self-Compton (SSC) scenario, if the redshift of the source is less than z$ \lesssim 0.4$.}


   \maketitle
%

\section{Introduction}
\label{sec:intro}

With only few exceptions, the Active Galactic Nuclei (AGN) so far detected in very high-energy $\gamma$-rays (VHE; E $\gtrsim$ 100\,GeV) belong to the class of blazars and are thought to have their jets closely aligned to the
line of sight, which leads to an amplification of their apparent luminosity through relativistic beaming. While flat-spectrum radio quasars (FSRQ) account for an important fraction of sources 
detected at energies in the MeV to GeV range \citep[$>$40\% of blazars of known type,][]{ack2011}, the population of blazars detected at VHE is largely dominated by BL 
Lac-type objects --- especially high-frequency peaked BL Lac objects (HBL), as defined by \citet{pad1995}.

A general difficulty in the study of BL Lac objects is that their redshifts are often unknown or not well determined, since spectral features from the host galaxy become undetectable due to the dominance of the non-thermal emission from the jet. Therefore, about 50\% of the BL Lac objects detected with {\it Fermi}-LAT \citep{ack2011} have an unknown redshift, and this is also the case for PKS\,0447-439. The source was first discovered in the radio band \citep{lar1981} and has since been detected in several observations in the radio, infrared, optical, ultraviolet and X-ray bands \citep{whi1994, gre1994, lam1997, cra1997, haa2009}. It was identified as a bright BL Lac object by \citet{per1998} and classified as an HBL by \citet{lan2008}, based on the ratio of the radio core luminosity at 1.4\,GHz over the X-ray luminosity at 1\,keV.

A first evaluation of its redshift of z$=$\,0.107 \citep{cra1997} was based on a mis-identification with a Seyfert 1 galaxy\footnote{private communication with H.~Landt and M.~V\'eron-Cetty}. Subsequent observations with the CTIO 4m telescope led to an estimation of z$=$0.205 \citep{per1998}. However, this claim was based only on a very weak spectral feature that the authors identifed as the Ca II line in an otherwise featureless spectrum. 
This result was not confirmed by a more recent attempt at determining the redshift of the source. The analysis by \citet{lan2008} yielded only a lower limit of z$>$0.176, based on the photometric method described by \citet{pir2007}, which was applied to a featureless spectrum resulting from observations with the CTIO and NTT telescopes. 

A recent analysis of observations of the source, which were carried out with the CTIO and NTT in 2007, puts a very high lower limit of z $\ge$\,1.246 on its redshift~\citep{lan2012}. This result is based on weak absorption lines interpreted as the Mg II $\lambda$2800 doublet. Such a high redshift is very difficult to reconcile with the detection of $\gamma$-rays from the source with energies of several TeV, given the absorption by the extragalactic background light (EBL). A more likely explanation is that the feature in the optical spectra corresponds to atmospheric absorption. The most 
recent data~\citep{pit2012}, taken with the X-Shooter telescope (ESO/VLT), do not confirm the high redshift value suggested by~\citet{lan2012}, but show instead a featureless spectrum apart from atmospheric absorption lines, at wavelengths that coincide exactly with the putative Mg II doublet suggested by ~\citet{lan2012}. This has also been confirmed subsequently by observations with the Clay Magellan II telescope~\citep{fum2012}.

While the optical spectrum of \pks is featureless and seems strongly dominated by non-thermal emission from the jet, observations with ATCA in the radio band show an extended, lobe-dominated source \citep{lan2008}. The emission from the jet, although strong at optical energies, is much weaker at radio energies, possibly due to synchrotron self-absorption.

High energy (HE) emission in the MeV to GeV range from the source has been detected with {\it Fermi}-LAT~\citep{atw2009}.  \pks is associated with 0FGL J0449.7-4348, one of the brightest blazars in the {\it Fermi} Bright Source List \citep{abd2009}, and has a hard spectrum in the HE band, which has made it a prime target for observations in the VHE range. In 2009, VHE emission from the source was discovered with the H.E.S.S. (High Energy Stereoscopic System) Cherenkov telescope array \citep{rau2009} and preliminary results were presented \citep{zec2011}.

In the present work, the final results of the H.E.S.S. analysis, using an updated analysis chain, are presented, together with simultaneous data at other wavelengths. Constraints on the emission processes and on the redshift of the source are discussed. Details on the VHE data analysis are provided in Section~\ref{sec:hess}. The discovery of VHE emission triggered observations of the source with the {\it RXTE} and {\it Swift} space telescopes in the X-ray, UV and optical bands. All the available multi-wavelength (MWL) data, including also optical data from the ATOM and ROTSE telescopes, situated on the H.E.S.S. site, are presented in Section~\ref{sec:mwl}. The spectral energy distribution (SED) and the temporal flux evolution in different energy bands are discussed in Section~\ref{sec:results}. The comparison of the spectra measured with {\it Fermi}-LAT and H.E.S.S. is used to derive a firm upper limit on the redshift of the source. An interpretation of the SED following a standard synchrotron self-Compton (SSC) scenario is suggested in Section~\ref{sec:modelling} for different assumptions on the redshift of the source.

Throughout the paper, a concordance cosmology model with $H_0 =$ 70\,km s$^{-1}$  Mpc$^{-1}$, $\Omega_\mathrm{M} =$ 0.3, and $\Omega_\mathrm{{\Lambda} }=$ 0.7 is assumed. Unless otherwise indicated, all errors are statistical and are given at the 1$\sigma$ (standard deviation) confidence level. 


\section{H.E.S.S. data analysis}
\label{sec:hess}

A total of 13.5 hours of good quality data were taken between November 2009 and January 2010 at the source position of \pks with the Cherenkov telescopes of the H.E.S.S. array, located in the Khomas highland in Namibia. Data taken when less than three of the four telescopes were fully operational have been rejected. Most of the data presented here are from December 2009. Only 0.9\,hr of observations from November 2009 and 0.4\,hr from January 2010 were kept in the final data set after standard quality selection \citep{aha2006}. Zenith angles of the observations range from 20$^{\circ}$ to 26$^{\circ}$.

Data were analysed using the \textit{model} analysis \citep{den2009}, which yields a more than twice as high significance for this source than a standard Hillas-type analysis~\citep{aha2006}, mainly due to a better rejection of the cosmic-ray background. For the given range of zenith angles and the minimum requirement of 60 photoelectrons per image used in the analysis, the analysis yields an energy threshold of about 220\,GeV. All results were cross-checked with independent analysis procedures and calibration chains \citep{aha2006}. The data yield a strong VHE signal at a statistical significance of 15.2$\sigma$.  The flux measured above 250 GeV was $\sim$3$\%$ of the flux from the Crab Nebula measured with H.E.S.S above the same threshold. The best fit position of the VHE $\gamma$-ray excess (\mbox{$\alpha_{J2000}=4^\mathrm{h}49^\mathrm{m}28.2^\mathrm{s} \pm 1.4^\mathrm{s}$ (stat) $\pm 1^\mathrm{s}$ (sys)}, \mbox{$\delta_{J2000}=-43^{\circ}50\mathrm{'} 12\mathrm{''} \pm 15\mathrm{''} $ (stat) $\pm 18^\mathrm{''}$ (sys)}) is compatible with the nominal position~\citep{ver2010} of the source (\mbox{$\alpha_{J2000}=4^\mathrm{h}49^\mathrm{m}24.7^\mathrm{s}$}, \mbox{$\delta_{J2000}=-43^{\circ}50\mathrm{'} 9\mathrm{''}$}). 

\begin{figure}[!h]
\centering
  \includegraphics[width=\columnwidth]{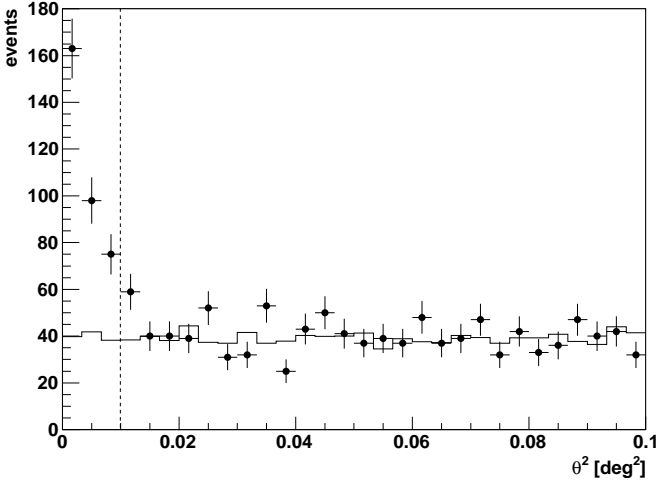}
  \caption{Distribution of the squared angular distance of $\gamma$-ray events from the nominal source position. The points represent the distribution for on-source events. The normalised off-source event distribution is described by the solid histogram. The vertical dashed line indicates the on-source integration region of 0.01 square degrees used in the analysis.}
  \label{fig:theta}
\end{figure}

The distribution of the $\gamma$-ray excess as a function of the squared angular distance from the nominal source position (Fig.~\ref{fig:theta}) has been determined using the ``reflected background'' method \citep{ber2007}.  From the on- and off-region, 336 and 1439 events were detected, respectively, with a relative normalisation factor between the regions of 0.083.The distribution of the on-source events is consistent with that expected from a point source.

An average power-law spectrum measured over the whole observation period has been extracted using forward folding with a maximum likelihood optimisation~\citep{pir2001}. The derived VHE spectrum is very soft, with a 
photon index $\Gamma =$ 3.89 $\pm$ 0.37 (stat) $\pm$ 0.22 (sys) and a flux normalisation at 1\,TeV of \mbox{$\Phi_{1 \mathrm{TeV}}=(3.5\pm1.1)\times10^{-13}\mathrm{cm}^{-2}\mathrm{s}^{-1}\mathrm{TeV}^{-1}$}. The equivalent $\chi^2$ of the fit is 36 for 26 degrees of freedom (d.o.f.) with a chance probability of 0.1. The systematic uncertainty in the flux normalisation is about 25$\%$. There is no significant indication for a spectral break or curvature in the observed spectrum. 

\begin{figure}[h!]
\centering
   \includegraphics[width=\columnwidth]{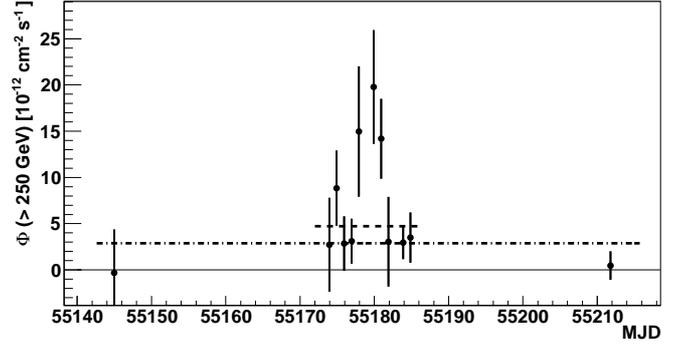}
  \caption{H.E.S.S. light curve from November 2009 to January 2010 in nightly bins with a constant fitted to the December data (dashed line)
  and to the whole dataset (dash-dotted line).}
  \label{fig:lc_hess}
\end{figure}

The nightly binned integrated flux above 250 GeV has been extracted using the measured photon index (cf. Table~\ref{tab:lc_hess} and Fig.~\ref{fig:lc_hess}). No significant signal is seen during the short observation time in November 2009 and January 2010. A fit of a constant to the nightly binned fluxes detected only during December 2009, weighted by the statistical uncertainties, yields an average flux level of \mbox{(4.7$\pm$1.0)$\times$10$^{-12}$ cm$^{-2}$ s$^{-1}$} with a $\chi^2$ of 16.3 for 9 d.o.f. and a chance probability of 0.06. 

When including the data from November and January in the fit (in nightly bins), a lower average flux of $(3.3\pm0.8) \times10^{-12}$ cm$^{-2}$ s$^{-1}$ with a $\chi^2$ of 22.2 for 11 d.o.f. (chance probability of 0.02) is found. The fractional RMS variability amplitude \citep{vau2003} over the complete observation period is F$_{\rm var}=$ 0.82$\pm$0.20. This is marginal evidence that the 
VHE flux was higher during December 2009 compared to the previous and the following month, although the statistics for these months are very limited due to the short duration of good-quality observations.

  \begin{table}[h!]
       $$
   \begin{array}{ p{0.2 \columnwidth}  p{0.2 \columnwidth} p{0.2 \columnwidth} p{0.3 \columnwidth} }
            \hline
             \noalign{\smallskip}
              t$_0$ & t$_1$ & $\Delta$t & $\Phi$($>$250 GeV) \\
            \noalign{\smallskip}
            \hline   
            \noalign{\smallskip} 
          55144.94  &	55144.98  &	 3090	 &	-0.32 $\pm$	4.15\\
        	55173.91 &	55173.95 &	 3245	&	$\,\,$2.73 $\pm$	4.51\\
        	55174.90 &	55174.96 &	 4836	&	$\,\,$8.83 $\pm$ 	3.75\\
        	55175.92 &	55175.96 &	 3248	&	$\,\,$2.85  $\pm$	2.63\\
        	55176.91 &	55176.95 &	 3234	&	$\,\,$3.10  	$\pm$2.21\\
        	55177.90 &	55177.95 &	 3255	&	$\,\,$14.97$\pm$	6.46\\
        	55179.87 &	55179.94 &	 4892	&	$\,\,$19.79$\pm$	5.73\\
        	55180.87 &	55180.93 &	 4846	&	$\,\,$14.18$\pm$	4.05\\
        	55181.86 &	55181.90 &	 3241	&	$\,\,$3.05  $\pm$	4.26\\
        	55183.83 &	55183.93 &	 8136	&	$\,\,$2.95  	$\pm$1.66\\
        	55184.83 &	55184.89 &	 4878	&	$\,\,$3.50  	$\pm$2.50\\
        	55211.80 &	55211.82 &	 1590	&	$\,\,$0.47  	$\pm$1.24\\
            \noalign{\smallskip}
            \hline
         \end{array} 
      $$
        \caption[]{Nightly averaged fluxes measured with the H.E.S.S. array. For each night, the start t$_0$ and end t$_1$ of observations (in MJD), the live time $\Delta$t (in seconds) and the integrated $\gamma$-ray flux above a threshold of 250 GeV (in units of 10$^{-12}$ cm$^{-2}$ s$^{-1}$) are shown.}
        \label{tab:lc_hess}
  \end{table}    

\begin{figure}
\centering
 \includegraphics[width=\columnwidth]{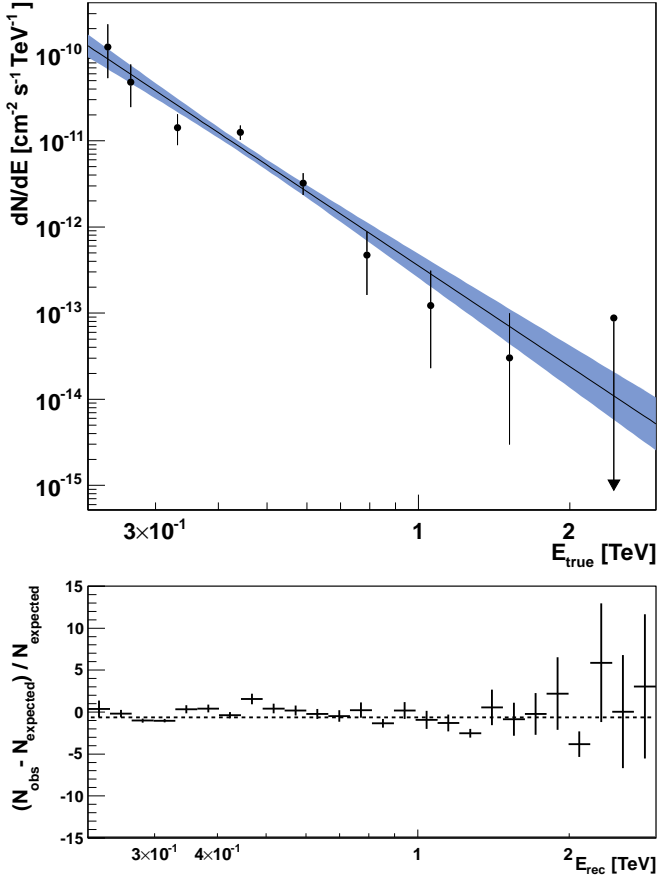}
  \caption{H.E.S.S. spectrum of the source extracted from the complete data set. The differential flux points (with an excess significance of 1$\sigma$) and the 68\% confidence band for the best fit are shown in the upper panel, as a function of true energy. The arrow represents an upper limit at the 99\% confidence level. The residuals of the maximum likelihood fit are given in the lower panel as a function of the reconstructed energy, together with a fit of a constant (dashed line).}
   \label{fig:spec}
\end{figure}


\section{Multiwavelength data}
\label{sec:mwl}

\subsection{High Energy data from {\it Fermi}-LAT}
\label{subsec:mwl_fermi}

\pks has been detected in all three {\it Fermi}-LAT point-source catalogues and is associated with the source 2FGL J0449.4-4350 \citep{ack2011} in the latest catalogue. 
A fit of a constant to the light curve from the {\it Fermi}-LAT two-year point-source catalogue yields an average photon flux between 0.1 GeV and 100 GeV of  (7.75 $\pm$ 0.24) $\times$ 10$^{-8}$ cm$^{-2}$ s$^{-1}$. The energy flux in this range is quoted as (1.36 $\pm$ 0.06) $\times$ 10$^{-10}$ erg cm$^{-2}$ s$^{-1}$. The photon index, derived from the likelihood analysis for 0.1 GeV to 100 GeV, has a value of 1.86 $\pm$ 0.02, indicating a hard spectrum. The source is seen to be variable (variability index of 91.9), with the maximum peak occurring on MJD 54757 (October 10, 2008). The monthly averaged flux around this peak between 0.1 GeV and 100 GeV is (1.56 $\pm$ 0.16) $\times$ 10$^{-7}$ cm$^{-2}$ s$^{-1}$.

{\it Fermi}-LAT data have been analysed between MJD 54682 (August 4, 2008) and MJD 55819 (September 15, 2011), in an energy range from 0.1 GeV to 200 GeV. The \fermi\ analysis was performed on the publicly available data, using the binned likelihood method \citep[cf.][]{ack2011} from the \textit{Science Tools} package, version \texttt{v9r23p1}\footnote{http://fermi.gsfc.nasa.gov/ssc/data/analysis/software/}, provided by the \fermi\ collaboration. Source class events are considered in a circular region of interest with a radius of 10\degr\ around the nominal position of PKS\,0447-439, using the \texttt{P7V6\_SOURCE} instrumental response functions. A cut  was applied on the zenith angle with respect to the Earth ($<$100$^{\circ}$), and on the rocking angle ($<$52$^{\circ}$). All the objects included in the 2FGL within 15\degr\ were included in the model reconstruction of PKS\,0447-439. The isotropic model \texttt{iso\_p7v6source} was used to account for both the extragalactic diffuse emission and residual instrumental background, while the spatial template \texttt{gal\_2yearp7v6\_v0} was used to account for the contribution from the Galactic diffuse emission.

\begin{figure}
\centering
 \includegraphics[width=1\columnwidth]{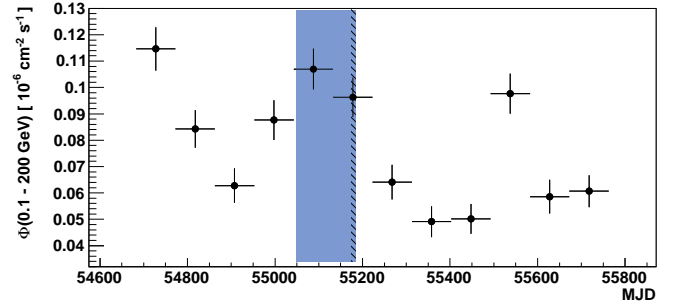}
  \caption{\fermi\ long-term lightcurve of \pks showing the photon flux integrated between 0.1 GeV and 200 GeV in intervals of 90 days. The horizontal error bars indicate the duration of the intervals. The period chosen for the extraction of the \fermi\ spectrum is indicated by the large blue band. The thin, striped band corresponds to the H.E.S.S. observational period in December 2009.}
   \label{fig:lc_fermi_longterm}
\end{figure}

This analysis yields a photon index of 1.86 $\pm$ 0.02 and an integrated energy flux between 0.1 GeV and 200 GeV of 
(1.45 $\pm$ 0.05) $\times$ 10$^{-10}$ erg cm$^{-2}$ s$^{-1}$. A long-term light curve with a binning of 90 days has been extracted, assuming a power-law spectrum with a fixed index of 1.86 (Fig.~\ref{fig:lc_fermi_longterm}). A second analysis has been carried out with a maximum energy of 100 GeV, as used for the 2FGL, yielding an integrated energy flux between 0.1 GeV and 100 GeV of (1.26 $\pm$ 0.05) $\times$ 10$^{-10}$ erg cm$^{-2}$ s$^{-1}$ and a photon index of 1.85 $\pm$ 0.02, in agreement with the results from the 2FGL. The estimated systematic uncertainty on the flux in these analyses is 10$\%$ at 0.1 GeV, 5$\%$ at 0.5 GeV and 10$\%$ at 10 GeV and above \citep[cf.][]{ack2011}.

In the \fermi\ long-term light curve (cf. Fig.~\ref{fig:lc_fermi_longterm}), a period of relatively constant flux was identified between August 7, 2009 (MJD 55050) and December 20, 2009 (MJD 55185), an interval which includes the nights of H.E.S.S. observations of the source in 2009. For this selected period, an averaged spectrum with a power-law shape was extracted with a photon index of 1.85 $\pm$ 0.05 (included in Fig.~\ref{fig:sed1}). Using the \textit{gtlike} tool and assuming a power-law shape for the source spectrum, the test statistic (TS) of the likelihood analysis of \pks is 1143, corresponding approximately to a 34$\sigma$ detection. The TS is 1143.43 for the power-law fit, 1143.66 for a log-parabolic fit, and 1146.23 for a fit with a broken power law. Comparing the log-parabolic model with the power law yields a log likelihood ratio (LLR) of 0.30. For the comparison between the broken power-law and the power-law model, the LLR is 2.96.
Thus, spectral shapes more complex than a power law do not result in an improvement of the likelihood for PKS\,0447$-$439. 

For the period of constant flux (MJD 55050--55185), an integrated flux between 0.1 GeV and 200 GeV of (1.08 $\pm$ 0.09) $\times$  10$^{-7}$ cm$^{-2}$ s$^{-1}$ was found. A fit of a constant to the selected period of the \fermi\ light curve (in bins of 3 day) yields a chance probability of 0.98 ($\chi^2 /$d.o.f. $=$ 27.3 / 44). The most energetic photon detected within the 95\% point-spread function (PSF) containment radius was found at 4.6 arcminutes from the nominal source position, with an energy of 95.5 GeV. For the full period (MJD 54682--55819), the most energetic photon within the PSF of the source has an energy of 147.5 GeV, while the energy resolution at this energy is about 12$\%$. A part of the \fermi\ light curve, binned in intervals of three days, is included in Fig.~\ref{fig:lc_mwl}.
 
\subsection{X-ray data from {\it RXTE} and {\it Swift} XRT}

The discovery of a VHE signal from \pks triggered observations with the {\it RXTE}  and {\it Swift} space telescopes. In December 2009, 11 pointings were taken with the PCA units of {\it RXTE} \citep{jah1996}. The STANDARD2 data were extracted using the {\tt HEASOFT 6.9} analysis software package provided by NASA/GSFC, and filtered using the {\it RXTE} Guest Observer Facility (GOF) recommended criteria. Only signals from the top layer of the PCA were selected. The faint background model was applied. Due to low statistics, the data from several consecutive pointings had to be added together to allow the extraction of energy spectra (cf. Table~\ref{tab:rxte}). The spectra were extracted between 3 and 12 keV using {\tt XSPEC v.12.6.0}, with a fixed column density of $N_{\rm H} = 1.24\times 10^{20}\,\rm cm^{-2}$ \citep{kal2005}. 

A power-law model was used for spectral fitting. No significant improvement was noted when using a broken power law. The parameters of the spectra for the different pointings are listed in Table~\ref{tab:rxte} and the light curve is included in Fig.~\ref{fig:lc_mwl} and will be discussed in Section~\ref{subsec:results_lc}. 

  \begin{table*}[h!]
       $$
         \begin{array}{l l l l l l l l}
            \hline
             \noalign{\smallskip}
            t_0 & t_1 & \Delta t & \mathrm{CR} & \rm{F}(2-10 \,{\rm keV}) & \Gamma &  \chi^2_r &  \mathrm{ndof} \\
            \noalign{\smallskip} \hline    \noalign{\smallskip}
55182.80 & 55182.83 & 1664 & 1.03 \pm 0.18 &1.33 \,$-$\, 0.29 + 0.06  & 2.77 \pm 0.29 & 0.97  & 19 \\  \hline 
\noalign{\smallskip}
55184.16 & 55184.20 & 1152 & 1.47 \pm 0.23 & \multirow{3}{*}{1.87 - 0.11 + 0.05 } & \multirow{3}{*}{3.05 $\pm$ 0.12} & \multirow{3}{*}{ 0.54} & \multirow{3}{*}{19} \\ 
\noalign{\smallskip}
55184.77 & 55184.79 & 1200 & 1.66 \pm 0.20  & &  &  &  \\ 
\noalign{\smallskip}
55184.83 & 55184.85 & 1104 & 1.54 \pm 0.22  & &   &  &  \\ \hline 
\noalign{\smallskip}
55184.90 & 55184.92 & 1072 & 2.61 \pm 0.31 &1.92\,$-$\,0.23 + 0.07 & 3.38 \pm 0.23 & 0.44 & 19  \\ \hline
\noalign{\smallskip}
55185.74 & 55185.77 & 1904 & 0.32 \pm 0.17 & \multirow{4}{*}{0.99 - 0.09 + 0.03} & \multirow{4}{*}{3.21 $\pm$ 0.17}  & \multirow{4}{*}{0.93} & \multirow{4}{*}{19}   \\  
\noalign{\smallskip}
55185.81 & 55185.83 & 1440 & 0.58 \pm 0.19 & &&   &   \\ 
\noalign{\smallskip}
55186.19 & 55186.20 & 1344 & 0.47 \pm 0.21 & &&  &  \\ 
\noalign{\smallskip}
55186.72 & 55186.75 & 1952 & 0.69 \pm 0.17 & & &   &   \\ \hline 
\noalign{\smallskip}
55186.79 & 55186.81 & 1536 & 1.92 \pm 0.25  & 0.94 \,$-$\, 0.26 + 0.05 & 2.90 \pm 0.29  & 0.44  & 19  \\ \hline 
\noalign{\smallskip}
55187.70 & 55187.73 & 2016 & 0.37 \pm 0.16 & 0.95 \,$-$\, 0.45 + 0.04 & 3.36  \pm 0.49  & 0.54  & 19 \\ 
\noalign{\smallskip}
          \hline
         \end{array} 
      $$
        \caption[]{Hard X-ray spectra measured with {\it RXTE}/PCA. The columns provide for each pointing the start t$_0$ and end t$_1$ of observations (in MJD), the live time $\Delta$t (in seconds), the count rate (per second), the integrated deabsorbed flux between 2 and 10 keV (in units of 10$^{-11}$ erg cm$^{-2}$ s$^{-1}$), the photon index, reduced $\chi^2$ and the number of degrees of freedom. Data from the observations in the second to fourth row and in the sixth to ninth row have been summed up to extract spectra with improved statistics.}
        \label{tab:rxte}
  \end{table*}    

In the soft X-ray band, the source was observed with {\it Swift} XRT \citep{bur2005} in 7 pointings taken in December 2009. The analysis of the XRT data (in photon-counting mode) was carried out with the {\tt HEASOFT 6.9} standard tools. Source events were extracted within a circle with a radius of 20 pixels ($\sim$47 arcseconds), while background events were extracted from both an annular region around the source (between 30 and 80 pixels) and a nearby source-free circle with a radius of 80 pixels. It was found that in some pointings pile-up was present. In these cases, the internal part of the PSF was excluded in the event selection and data were extracted in an annular region with an outer diameter of 30 pixels, while the inner radius was chosen by fitting the radial distribution with a King profile. Using the \textit{grppha} tool, bins were grouped to yield at least 30 counts per bin, and the data were fitted with a single or broken power law, with the same fixed column density as given above. Only points in the intervals from 0.3 to 0.45 keV and from 0.6 to up to 6 keV were included in the fit to suppress known systematic effects at intermediate energies \citep{cam2006}. 

The results from the power-law and broken power-law fits were compared using an F-test, where a broken power law with a break at around 1 keV  was preferred at a probability of $>$99\% for two pointings, as indicated in Table~\ref{tab:xrt}. The hypothesis of additional, intrinsic absorption was also tested and was not found preferable over the power laws with Galactic absorption.
The light curve of the soft X-ray flux is included in Fig.~\ref{fig:lc_mwl} and will be discussed in Section~\ref{subsec:results_lc}.  

 \begin{table*}[h!]
       $$
         \begin{array}{l l l l l l l l l l l l}
            \hline
             \noalign{\smallskip}
             t_0 & t_1 & \delta t & CR & F(0.3-4 \,{\rm keV}) & \Gamma_1 & E_{break} & \Gamma_2 &  \chi^2_r & \mathrm{ndof} \\
            \noalign{\smallskip}  \hline  \noalign{\smallskip}
55181.85 & 55182.12 & 2825 & 0.78 \pm 0.02 & 2.42 - 0.20 + 0.04 & 2.02 \pm 0.18 & 0.79 \pm 0.15  & 2.56 \pm 0.07 & 0.93 & 47 \\ 
\noalign{\smallskip}
55183.11 & 55183.12 & 982  & 1.37 \pm 0.04 & 4.33 - 0.13 + 0.11  &  2.31 \pm 0.05 & -  & - & 0.93 & 30 \\ 
\noalign{\smallskip}
55183.18 & 55183.32 & 2795 &  1.14 \pm 0.02 & 3.71 - 0.11 + 0.05 & 2.10  \pm 0.08  & 1.11  \pm 0.24 & 2.43 \pm 0.08 & 0.92 & 69 \\ 
\noalign{\smallskip}
55184.10 & 55184.12 & 1515 & 0.76 \pm 0.02 & 3.44 - 0.14 + 0.10  & 2.19  \pm 0.06 & - & - & 1.21 & 26 \\ 
\noalign{\smallskip}
55184.17 & 55184.25 & 3106 &  0.79 \pm 0.02 & 2.95 - 0.07 + 0.07  & 2.32 \pm 0.04 & -  & - & 1.04  & 51 \\ 
\noalign{\smallskip}
55190.00 & 55190.02 & 1281 & 0.69 \pm 0.02 & 2.60 - 0.13 + 0.14 & 2.42 \pm 0.07  & - & - & 0.50 & 18 \\ 
\noalign{\smallskip}
55190.07 & 55190.27 & 4769 & 0.70 \pm 0.01 & 2.53 - 0.06 + 0.06 & 2.49 \pm 0.03  & -  & - & 1.02 & 67 \\          
            \noalign{\smallskip}  \hline
    \end{array} 
      $$
        \caption[]{Soft X-ray spectra measured with {\it Swift} XRT. The columns provide for each night the start t$_0$ and end t$_1$ of observations (in MJD), the live time $\delta$t (in seconds), the count rate (per second), the integrated deabsorbed flux between 0.3 and 4 keV (in 10$^{-11}$ erg cm$^{-2}$ s$^{-1}$), the photon index below the break energy, the break energy (in keV), the photon index above the break energy, the reduced $\chi^2$ and the number of degrees of freedom.
        For nights where no break energy is given, $\Gamma_1$ is the photon index for a power law over the whole energy range.}
       \label{tab:xrt}
      \end{table*}    

The data taken with {\it Swift} XRT and {\it RXTE} PCA are not simultaneous, with the exception of MJD 55184 (Dec. 19, 2009), where an overlap of about 0.7 ks occurs. For this
night, data from the first pointing with PCA have been jointly analysed with data from the second pointing with XRT. A fit of the combined spectra with a power law yields a photon
index of 2.31 $\pm$ 0.03 and a deabsorbed energy flux of ($1.16_{- 0.11}^ {+ 0.06}$) $\times$ 10$^{-11}$ erg cm$^{-2}$ s$^{-1}$ between 2 and 10 keV for a fixed $N_{\rm H} = 1.24\times 10^{20}\,\rm cm^{-2}$. This fit has a reduced $\chi^{2}$ of 0.91 for 72 d.o.f. A fit with a broken power law does not yield a significant improvement.

\subsection{Ultra-violet and optical data from {\it Swift} UVOT, ATOM and ROTSE}

Simultaneously to the {\it Swift} XRT pointings, observations in the ultra-violet (UV) and optical bands were made with the {\it Swift} UVOT instrument \citep{rom2005}. Data were taken with the six available filters (v, b, u, UVW1, UVM2, UVW2 - in order of increasing frequency). UVOT counts were extracted within a radius of 5 arcseconds centered on the source. Flux densities have been extracted with
\textit{uvotmaghist v1.1} and have been multiplied with the central wavelength of each filter \citep{poo2008}. Galactic extinction has been corrected for, following \citet{rom2009}, with an extinction coefficient E$_{\rm B-V} =$ 0.014 \citep{sch1998}. The light curves are included in Fig.~\ref{fig:lc_mwl} and nightly averaged fluxes are listed in Table~\ref{tab:uvot}.

  \begin{table*}[h!]
       $$
         \begin{array}{l l l l l l l}
            \hline
             \noalign{\smallskip}
            t & F_{v} & F_{b} & F_{u} & F_{UVW1} & F_{UVM2} & F_{UVW2}  \\  
            \noalign{\smallskip} \hline    \noalign{\smallskip}
            55182 &  4.81 \pm 0.12* &  5.16 \pm 0.14* & 5.13 \pm 0.52* & 5.02 \pm 0.59* & 5.68 \pm 0.19* & 5.04 \pm 0.19* \\
            \noalign{\smallskip}
            55183 & 4.99 \pm 0.15 & 5.37 \pm 0.13 & 5.38 \pm 0.142 & 5.43 \pm 0.02* &  5.95 \pm 0.15 & 5.23 \pm 0.12 \\
            \noalign{\smallskip}
            55184 & 4.43 \pm 0.13 &  5.04 \pm 0.11 & 5.04 \pm 0.04* & 5.05 \pm 0.11 & 5.50 \pm 0.13 &  4.93 \pm 0.11 \\
            \noalign{\smallskip}
            55190 & 4.38 \pm 0.12 & 4.92 \pm 0.11 & 4.98 \pm 0.12 & 4.98 \pm 0.11 & 5.41 \pm 0.09* & 4.88 \pm 0.10 \\
 \noalign{\smallskip}   
          \noalign{\smallskip}  \hline
         \end{array} 
      $$
        \caption[]{Nightly averaged fluxes in the optical and ultra-violet bands measured with {\it SWIFT} UVOT. The date of the nights is given in the first column (in MJD).
                The remaining columns provide the deabsorbed flux densities for the different filters (v, b, u, UVW1, UVM2, UVW2), multiplied with the central wavelengths
                 (5468 $\AA$, 4392 $\AA$, 3465 $\AA$, 2600 $\AA$, 2246 $\AA$, 1928 $\AA$) in units of 10$^{-11}$ erg cm$^{-2}$ s$^{-1}$. Errors represent the RMS
                  spread between points for nights where more than one measurement was available (marked with an asterisk *) and the statistical uncertainty in the single measurement otherwise. }
        \label{tab:uvot}
  \end{table*}

Apart from the data from the {\it Swift} UVOT instrument, optical data are also available over longer intervals from the ground-based ATOM \citep{hau2004} and ROTSE-IIIc \citep{ake2003} telescopes, both located on the H.E.S.S. site. 

Data from the 75 cm ATOM telescope are available in the B and R bands from September 2009 to December 2012. During the H.E.S.S. campaign, only few observations were taken  in the B band and some of them were of insufficient quality to yield acceptable flux measurements due to the presence of clouds. A 4 arcsecond radius of aperture was used to extract the data. In the R band, the ATOM light curve (Fig.~\ref{fig:lc_mwl}) shows a slow flux decrease in December 2009. In the 2009 ATOM data and up to mid 2010, the magnitude increased by about 0.7 mag. The source was still in a relatively high flux state in December 2009 as compared to the following months. On longer time scales, variations of about 1 mag were observed, and the source reached again about the same flux level as in December 2009 by the end of 2010 and by the end of 2011.

ROTSE has a wide field of view (1.85$^{\circ}~\times$1.85$^{\circ}$) and is operated without filters. A relative R magnitude is derived by comparison of the instrumental magnitude with the USNO catalogue~\citep{ake2000}. The ROTSE light curve, included in Fig.~\ref{fig:lc_mwl}, shows the same slow flux decrease during December 2009 as seen in the ATOM data. Long-term optical data taken with ROTSE between 2005 and 2010 confirm that the optical flux was decreasing, but in a relatively high state in December 2009.

The data from both instruments have been corrected for Galactic absorption as described above. 


\section{Results}
\label{sec:results}

\subsection{Flux evolution in different wavebands}

\label{subsec:results_lc}

 \begin{figure*}
     \centering
      \includegraphics[width=19cm]{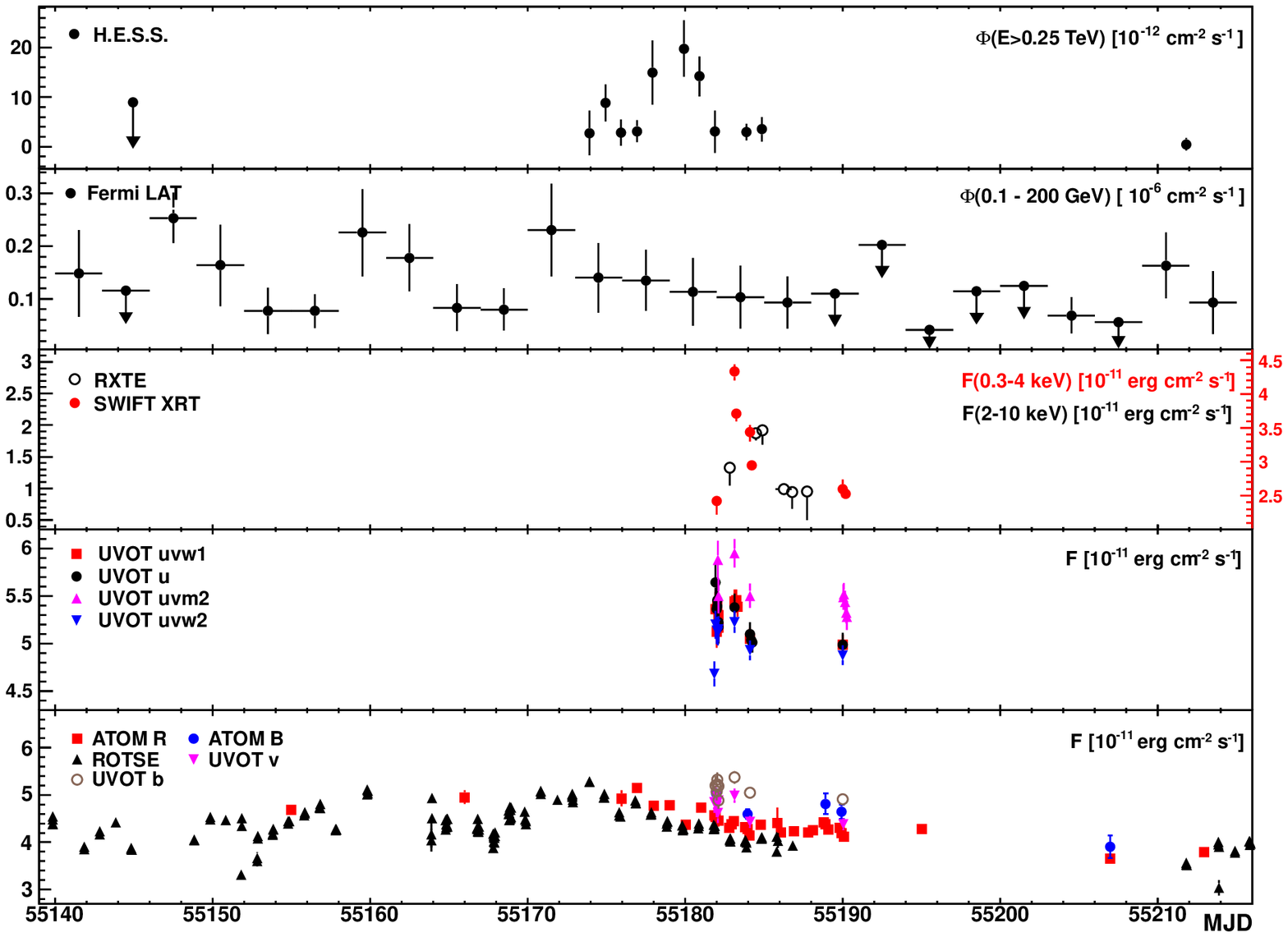}
  \caption{Multi-wavelength light curves for the time period from November 2009 to January 2010. The first panel shows the integrated flux above 250 GeV seen with H.E.S.S., while the {\it Fermi}-LAT flux between 0.1 and 200 GeV is presented in the second panel. Horizontal error bars indicate the duration of the
  time intervals over which fluxes were averaged. Upper limits are given at the 95\% confidence level (CL) in the H.E.S.S. and {\it Fermi} data. The third panel shows the X-ray fluxes between 2 and 10 keV from {\it RXTE} (open circles, y-axis on the left) and between 0.3 and 4 keV from {\it Swift} XRT (filled circles, y-axis on the right). UV data taken with {\it Swift} UVOT with the u, UVW1, UVW2 and UVM2 filters are shown in the fourth panel. The last panel includes optical data from the b and v filters from UVOT, in the B and R bands from ATOM and in the R band from ROTSE. For the UV and optical data from UVOT, flux densities have been multiplied by the central wavelengths of the corresponding filters.}
  \label{fig:lc_mwl}
\end{figure*}

 \begin{figure}
     \centering
      \includegraphics[width=\columnwidth]{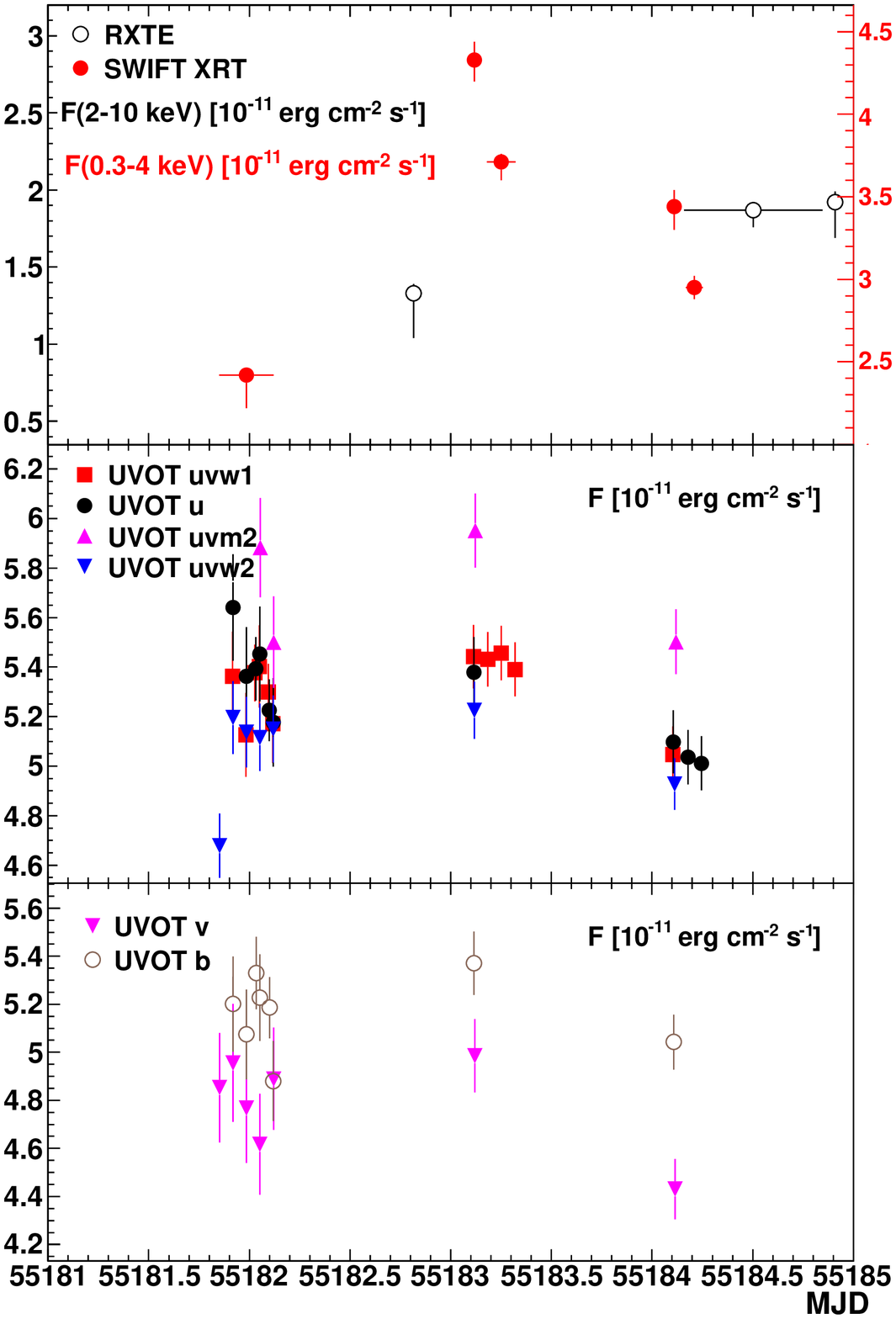}
  \caption{Zoom on the {\it RXTE}/PCA, XRT and UVOT light curves around the flare observed with XRT. Horizontal error bars indicate the
  duration of the time intervals over which fluxes were averaged. A flux decrease can be seen in the UVOT UV data that accompanies the
  decline in the X-ray flux after the flare.}
  \label{fig:lc_mwl_zoom}
\end{figure}

Figure~\ref{fig:lc_mwl} shows a comparison of the flux evolution in different energy bands during the time period where data from H.E.S.S. are available (November 2009 to January 2010). The first panel shows the H.E.S.S. light curve above 250 GeV in nightly intervals, with negative flux estimates replaced by upper limits at the 95\% confidence level following \citet{fel1998}. Details on the flux evolution in the H.E.S.S. data were given in Section~\ref{sec:hess}.

Over the same period, the {\it Fermi}-LAT light curve between 0.1 and 200 GeV, binned in intervals of three days, is shown in the second panel. (A light curve with one-day binning suffers from very large statistical uncertainties.)
The fractional RMS variability amplitude F$_{\rm var}=$ 0.21 $\pm$ 0.22 does not point to any significant intrinsic variability during this period. However, the fit of a constant to the light curve has a $\chi^2$ of 75.3 for 24 d.o.f., corresponding to a chance probability of less than 10$^{-6}$, which indicates a deviation from a constant flux after the stable period between MJD 55050 and MJD 55185 (cf. Sec.~\ref{subsec:mwl_fermi}), notably due to the low fluxes on MJD 55195 and MJD 55207.  
When comparing the H.E.S.S. and {\it Fermi}-LAT light curves, it can be seen that the absence of a positive excess in the H.E.S.S. data in November 2009 is reflected in the absence of a signal in the {\it Fermi}-LAT flux for the same night where the H.E.S.S. data were taken.

Rapid variability on nightly time scales is clearly detected in the X-ray data (third panel). Fits to a constant of the {\it Swift} XRT and {\it RXTE} PCA light curves yield chance probabilities below 10$^{-15}$. The fractional RMS variability amplitude for the data from XRT is F$_{\rm var}=$0.22 $\pm$ 0.01 and for PCA F$_{\rm var}=$0.32 $\pm$ 0.05.  In the hard X-ray band (2-10 keV), data from {\it RXTE} PCA show a flare on MJD 55185 (December 20). The flux drops to about half its peak value on the following night. As far as can be concluded from the limited statistics available, the flare is not accompanied by any significant spectral variation. In the soft X-ray band (0.3-4 keV), the {\it Swift} XRT light curve exhibits a flare on MJD 55183 (December 18, 2009), where the flux reaches a level about twice as high as its lowest value in this observation period. No significant correlation is found between the photon indices and the integrated fluxes for the different pointings. The minimum variability time scale in the X-ray bands has been estimated in terms of the flux doubling time T$_2$, following \citet{zha1999}. The smallest flux doubling time was found in the XRT light curve with a value of T$_2 = 0.87 \pm 0.24$ days (after discarding results with uncertainties above 30 $\%$), indicating flux variations on the order of one day.

It is important to note that none of the pointings of the two X-ray telescopes were exactly simultaneous, with some overlap occurring only during one pointing on MJD 55184 (see Fig.~\ref{fig:lc_mwl_zoom}). The flares seen in the two X-ray bands might be part of the same event, but this cannot be concluded with certainty. No corresponding flare can be seen in the {\it Fermi}-LAT  data, but flux variations of the order seen in the X-ray band might well be hidden in the statistical uncertainties. In the VHE band, no data are available during the night of the {\it Swift} XRT flare.

The data taken with {\it Swift} UVOT in the UV band (fourth panel of Fig.~\ref{fig:lc_mwl}) show significant variation in the u- and UVW1-filters, when fit with a constant flux (chance probabilities $<$ 10$^{-11}$; F$_{\rm var}=$ 0.08 $\pm$ 0.01 for the U-filter and F$_{\rm var}=$ 0.09 $\pm$ 0.01 for the UVW1-filter). Variation in the UVW2- and UVM2-filters is only marginally significant. A variation is seen in the flux evolution that is simultaneous to the flare detected with {\it Swift} XRT. When looking at the nightly averaged UV fluxes, the flux increase in the UVOT data between MJD 55182 and MJD 55183 is not significant, but there is evidence of a flux decrease of about 10$\%$ between MJD 55183, where the X-ray flare is seen, and the following night (see Fig.~\ref{fig:lc_mwl_zoom} and Table~\ref{tab:uvot}).

In the optical band (last panel of Fig.~\ref{fig:lc_mwl}), flux variations during the period where the X-ray flares are seen are not statistically significant (with a 2\% chance probability for a fit of a constant to the light curve in the UVOT v-filter), but there is an indication for a flux decrease between MJD 55183 and the following night in the v- and b-filters observed with {\it Swift} UVOT. This is not visible in the R band, covered with ATOM and ROTSE. ATOM data in the B band suffer from large uncertainties during several nights, since these observations are affected by clouds, and no significant variation was detected. 

\subsection{Spectral energy distribution}

\begin{figure*}
\begin{center}
  \includegraphics[width=18.5cm]{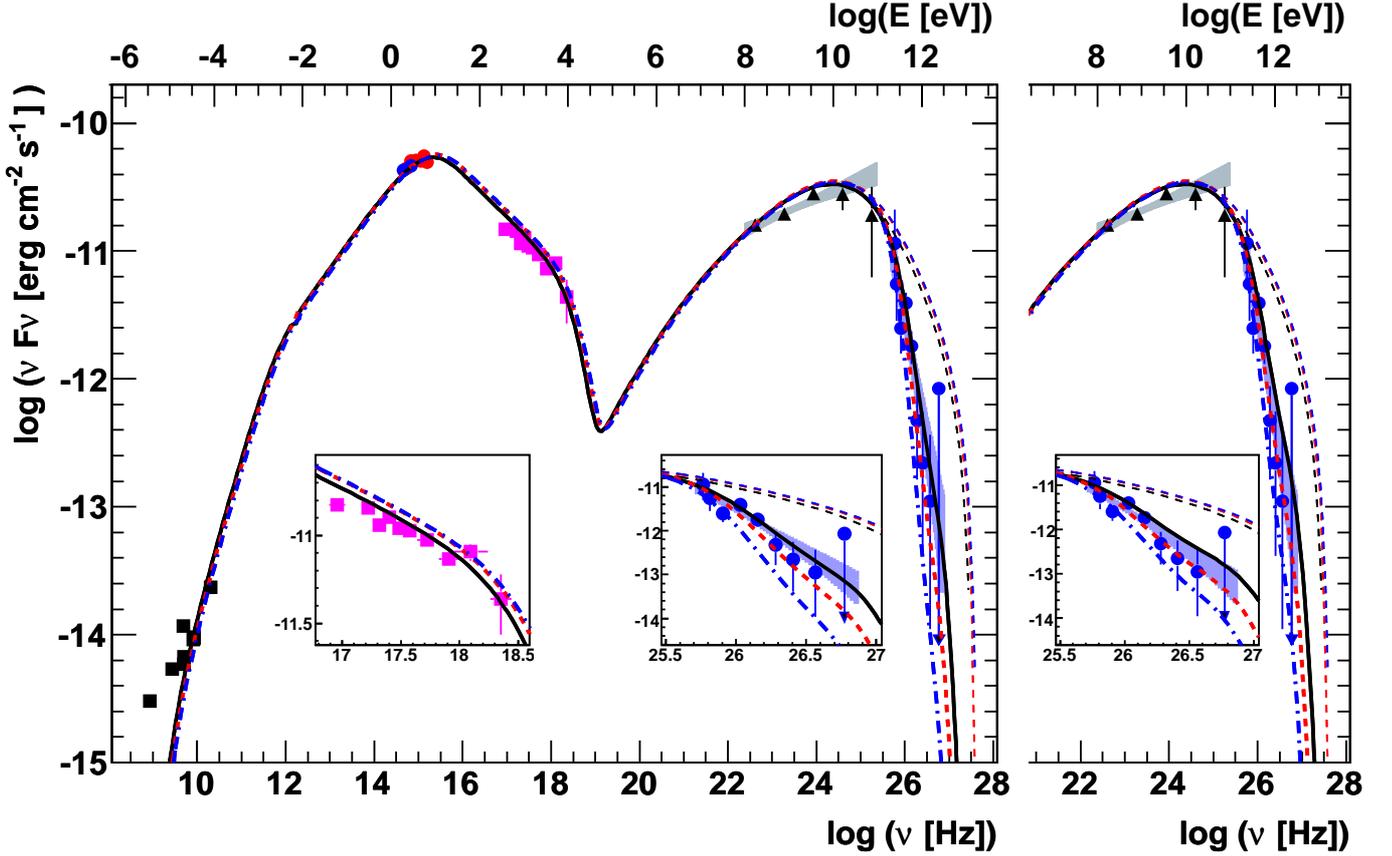}
  \end{center}
  \caption{Spectral energy distribution of \pks showing archival radio data from NED and SIMBAD/VizieR (black squares; details are given in the text), optical and UV data from ATOM (blue filled circles) and {\it Swift} UVOT (red filled circles), X-ray data from {\it Swift} XRT and {\it RXTE} PCA (magenta squares), $\gamma$-ray data from {\it Fermi}-LAT (black triangles and grey bow-tie) and VHE $\gamma$-ray data from H.E.S.S. (blue filled circles and light blue bow-tie). The models shown here correspond to a redshift of 0.2 (black solid line; deabsorbed spectrum shown by thin black dashed line), 0.3 (red
  dashed line; deabsorbed spectrum shown by thin red dashed line) and 0.4 (blue dash-dotted line; deabsorbed spectrum shown by thin blue dashed line). For the graph on the left panel, the EBL model uses the description by \citet{fra2008}. On the right panel, only the high-energy part of the SED is shown, with an EBL model following  the description by \citet{kne2010}, which presents a very low EBL level. The inlays show a zoom on the X-ray range and VHE range of the SED.}
  \label{fig:sed1}
\end{figure*}

The spectral energy distribution of \pks is presented in Fig.~\ref{fig:sed1} (the models are discussed in Section~\ref{sec:modelling}).
The VHE spectrum is not corrected for absorption by the EBL, since it depends on the unknown redshift of the source. The spectral slope is very steep, which might be at least partially due to absorption on the EBL.

In the high-energy $\gamma$-ray band, the rather flat {\it Fermi}-LAT spectrum, extracted over a period of more than four months of relatively constant flux (cf. Section~\ref{subsec:mwl_fermi}), suggests a wide peak, followed by the steep spectrum in the VHE range. 

In the X-ray range, the combined spectrum including {\it Swift} XRT and {\it RXTE} PCA data for MJD 55184 is included in the SED. These data represent an average flux level between the observed peak and the lowest observed flux in the {\it Swift} XRT light curve. {\it Swift} UVOT data and optical data from the ATOM telescope are also included for this night. 
The optical and X-ray data have been corrected for Galactic absorption as described in Section~\ref{subsec:results_lc}. Archival radio data from several catalogues
(PKSCAT90, \citet{wri1990}; PMN, \citet{wri1994}; SUMSS, \citet{mau2003}; CRATES, \citet{hea2007};  AT20G, \citet{mur2010}; ATPMN, \citet{mcc2012}) have been added to the SED.

The SED clearly shows the double-bumped structure that is characteristic of BL Lac objects. The peak frequency of the high-energy bump is relatively well constrained with the information from the {\it Fermi}-LAT and H.E.S.S. data, if a certain 
redshift of the source and EBL model are assumed. The detection of a flare in X-rays and a correlated variation in the UV flux suggests a common origin for the emission in these wavelength bands (at least above the R band).  Given the high flux level in the optical and UV range --- as compared to the lower X-ray flux --- and the rather hard X-ray spectrum, the synchrotron peak must be located not far above or within the UV range. It is possible that at the relatively long wavelengths seen in the R band, another component dominates the optical flux. There might be a contribution from the host galaxy, but given the featureless optical spectrum and the non-detection of an extension of the source so far, a non-thermal component seems more likely. An extended jet might be responsible for the emission at low energies, while the emission at higher energies could stem from a more compact region inside the jet. This would explain that variability is seen in the X-ray and UV range, but not (significantly) in the optical bands. The existence of different radiative components in the optical and high-energy bands in blazars is frequently found (see \citet{abr2012} for a recent example). 

\subsection{Upper limit on the redshift of PKS~0447-439}
\label{subsec:zlimit}

The combined {\it Fermi}-LAT and H.E.S.S. data set is used to derive an upper limit on the redshift of the source. No assumptions are made on the source emission characteristics, other than excluding spectral upturns beyond the {\it Fermi}-LAT energy band, which are not expected in general and would be theoretically difficult to account for. 

One case of a spectral upturn at high energies might have been seen during a flare of Mrk 501~\citep{ner2011}. \citet{lef2011} interpret this hypothetical spectral upturn as resulting from the combination
of several emission regions in a leptonic scenario with Maxwell-like particle disbributions. The observational evidence for this feature, however, is weak. For \pks, there is no indication of an upturn in the {\it Fermi}-LAT data.

Spectral upturns might also occur in a scenario where the $\gamma$-ray flux in the {\it Fermi}-LAT band suffers from internal absorption in the source, e.g. on emission from the Broad Line Region (BLR) \citep{sen2011, pou2012} or from the accretion disk or torus \citep[e.g.][]{don2003}, and recovers in the VHE band \citep[see also][]{zac2011}. However, the absence of line features in the optical spectra from BL Lac objects in general suggests that emission from the BLR is negligible in these sources. In the case of PKS\,0447-439, the available optical spectra, as discussed in Section~\ref{sec:intro}, do not show any significant features, except for absorption lines from the Earth's atmosphere. The $\gamma$-ray data from \pks are best described by power laws, with no indication for absorption on internal photon fields.

The {\it Fermi}-LAT power-law spectrum, constructed over the integration period between MJD\,55050 to MJD\,55185, is extrapolated to the highest energies to serve as an upper limit on the unabsorbed flux in the VHE band. Following the method proposed by \citet{geo2010}, the ratio between the VHE flux and the extrapolated {\it Fermi}-LAT flux provides an upper limit on the optical depth, as a function of energy, due to EBL attenuation. A comparison of this upper limit with the predictions of a given EBL model as a function of redshift can be translated into an upper limit on the redshift of the source. 
This method has been used to constrain the density of the EBL with data from {\it Fermi}-LAT and MAGIC from the FSRQ PKS\,1222+21 \citep{ale2011}. A similar approach was applied to {\it Fermi}-LAT and VHE data from the BL Lac object PG\,1553+113 to derive an upper limit on its unknown redshift \citep{abd2010}.

Intrinsic softening of the spectrum beyond the {\it Fermi}-LAT band, which might well occur in PKS\,0447-439, would only lead to a lower value for the inferred redshift and would thus not be in contradiction with the derived upper limit. 
A problem in the interpretation of the upper limit would only arise in non-standard emission scenarios, in which absorption on the EBL is partially avoided. This is the case for scenarios \citep[e.g.][]{ess2011, mur2012, aha2012, tay2011, aha2002}, where the observed $\gamma$-rays are interpreted as secondary photons produced outside of the source in interactions of primary photons or protons
with the cosmological microwave background and the EBL. Another possibility is given by scenarios involving exotic particles or interactions, such as photon-axion oscillations \citep{dea2011}. 

Modifying Equation~\ref{eq:tau} from \citet{ale2011} to include a correction term for the systematic error on the H.E.S.S. flux, the upper limit on the optical depth $\tau_{\rm max}$ is given by
\begin{equation}
\tau_\mathrm{{max}}(E) ~=~ \ln \left[ \frac{F_\mathrm{{int}}(E)} {(1 - \alpha) \cdot ( F_\mathrm{{obs}}(E) - 1.64 \cdot  \Delta F_\mathrm{{obs}}(E))}  \right] \, .
\label{eq:tau}
\end{equation}
Here $F_\mathrm{{int}}(E)$ is the intrinsic flux, assumed to be a simple extrapolation of the flux measured with {\it Fermi}-LAT, $F_\mathrm{{obs}}(E)$ is the VHE flux measured with H.E.S.S., with a statistical uncertainty of $\Delta F_\mathrm{{obs}}(E)$, and $(1-\alpha)$ is the factor accounting for the systematic error on $F_\mathrm{{obs}}(E)$. The systematic error is a measure of the uncertainty in the flux and is expressed as a fraction of the measured flux. For the analysis cuts used here and for the measured spectral index, the factor $\alpha$ equals about 0.26, as determined from detailed Monte Carlo simulations of the instrumental performance, corresponding to a $\sim26\%$ systematic error on the flux averaged between 0.3\, TeV and 2\, TeV. This is taken into account in a conservative way, in the sense that $\alpha$ is the maximum factor by which the flux could be overestimated. 

 \begin{figure*}[h!]
     \centering
      \includegraphics[width=9cm]{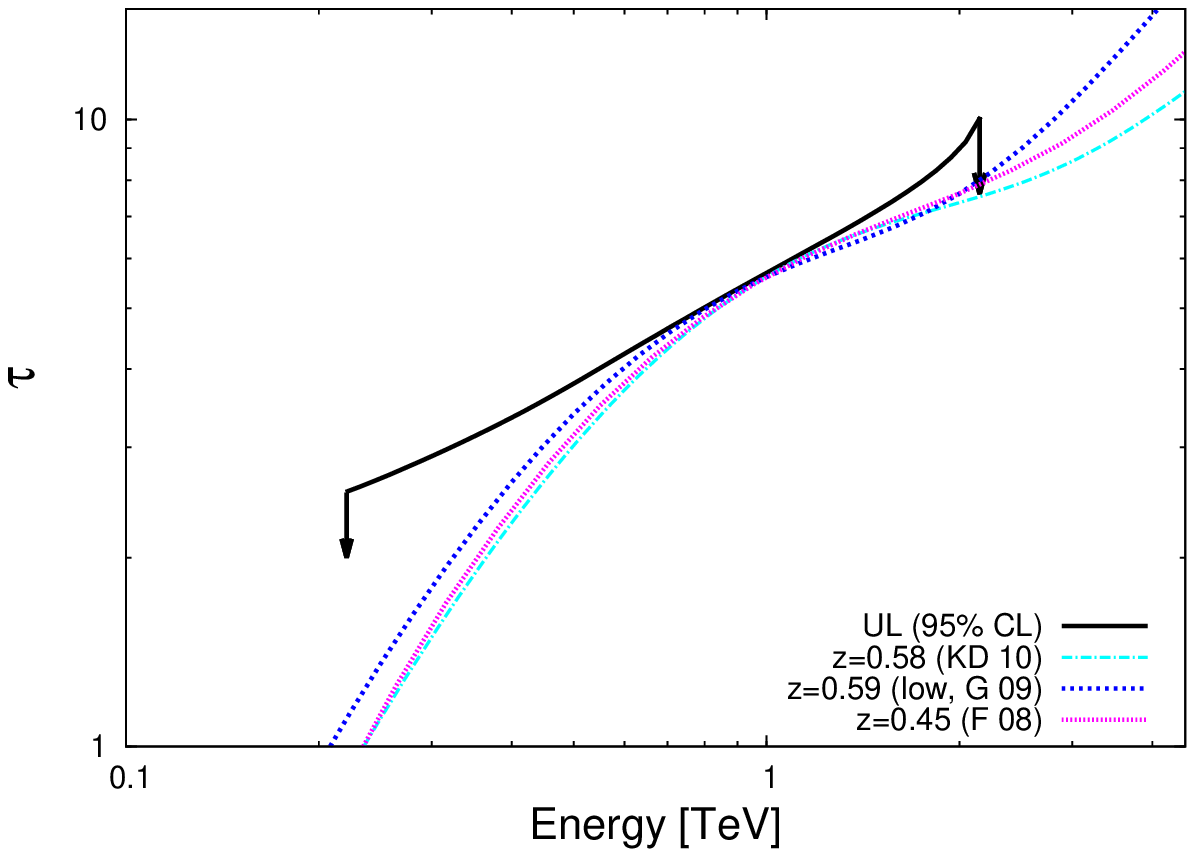}
      \includegraphics[width=9cm]{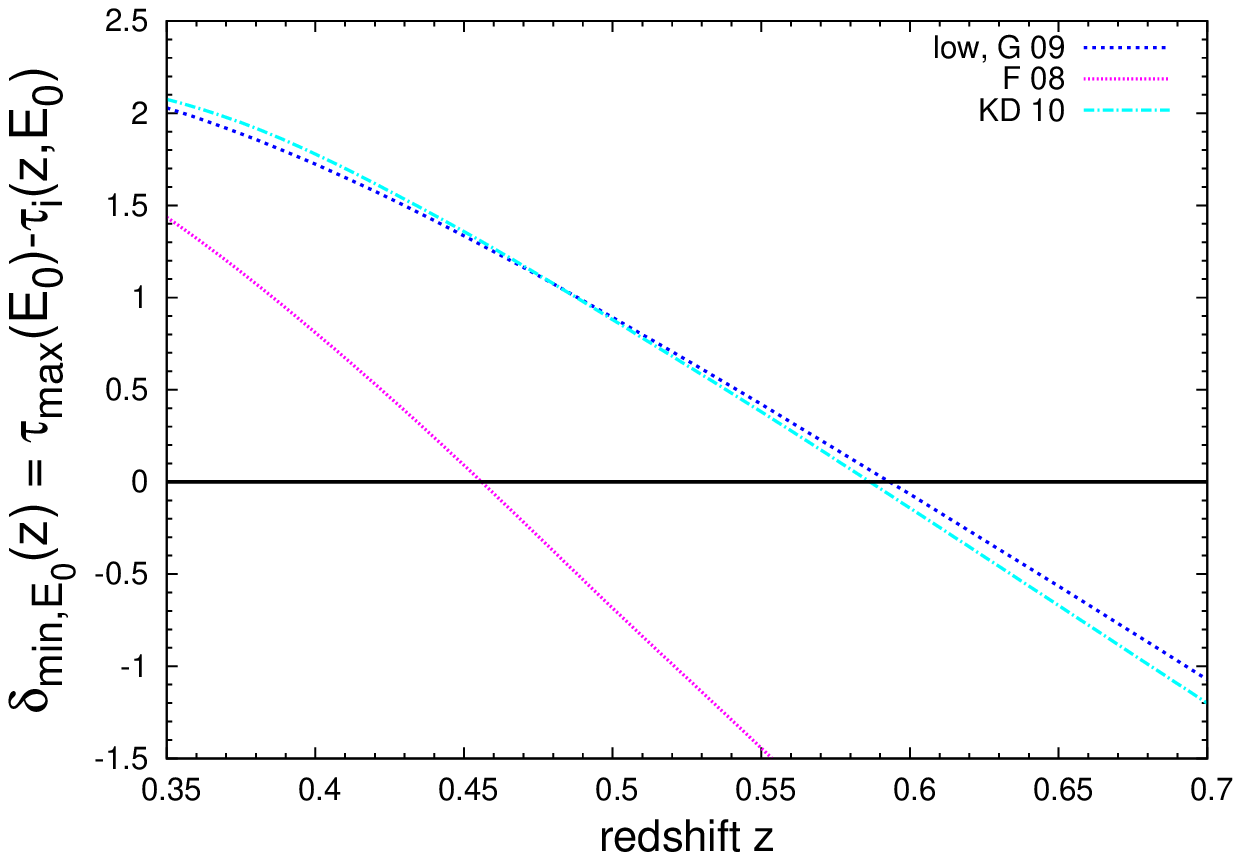}
        \caption{Left panel: upper limit at 95 \% CL on the optical depth as a function of energy (solid line) as derived from Equation~\ref{eq:tau}. The optical depth distribution derived with different EBL models at specific redshifts is shown for comparison (dash-dotted line: Kneiske and Dole 2010; dashed line: Gilmore et al. 2009, ``low'' model; dotted line: Franceschini et al. 2008. 
Right panel: the minimum difference $\delta_{\rm min,E_0}$ between the maximum optical depth derived here and the optical depths predicted by the same EBL models is shown as a function of the redshift. E$_0$ is the energy for which, at a given redshift, the difference between the derived and predicted optical depths is minimized. The upper limit on the redshift, for each EBL model under consideration, corresponds to $\delta_{\rm min,E_0} =$0.}
  \label{fig:zlimit1}
    \end{figure*}

As shown in Fig.~\ref{fig:zlimit1}, an upper limit of z$<$0.45 can be put on the redshift of the source at the 95$\%$ confidence level (CL), when using the EBL model by \citet{fra2008}, which generally provides a good description of the observed VHE spectra. Applying an estimation of the lower limit of the EBL level, as provided by the ``low'' model by \citet{gil2009} and by the model by \citet{kne2010}, a more conservative upper limit of z$<$0.59 and z$<$0.58 is derived, respectively. 

A detailed discussion of the inclusion of statistical and systematic uncertainties of the {\it Fermi}-LAT spectrum in the upper limit, not included in Equation~\ref{eq:tau}, is given in Appendix~\ref{app:stats}. It is shown that this approximation is indeed justified in the present case. 

An estimate of the redshift of \pks of z=\,0.20$\pm$0.05 has recently been suggested by \citet{pra2011}, based on preliminary data from H.E.S.S.~\citep{zec2011} and on an empirical relation between observed photon indices in the {\it Fermi}-LAT and VHE band \citep{pra2010}. Such an estimate is based on the assumption that all detected VHE blazars share very similar intrinsic spectral properties. The most conservative upper limit of z$<$0.59 derived in the present work, on the other hand, provides a firm
limit on the redshift of the source, which takes into account statistical and systematic uncertainties in the spectral measurements and depends only on minimal assumptions
on the spectral shape and the validity of current EBL models.


\section{Interpretation with an SSC Model}
\label{sec:modelling}

The interpretation of the multiwavelength spectrum of \pks is complicated by the fact that the redshift of the source is unknown, but it is possible to study the variation of the model parameters for different redshift values within the derived lower and upper limits. The modelling presented here, based on a simple SSC model, aims at investigating two issues: the physical parameters characterizing the emission region and the maximum allowable redshift of the source that can be accommodated under the assumption of a simple SSC scenario.

The stationary homogeneous one-zone SSC model used for this study is based on the code by \cite{Kat2001}. Synchrotron and SSC emission stem from a single population of relativistic electrons and positrons in a spherical plasma blob inside the jet. This blob is characterized by its radius $R$, Doppler factor $\delta$ and homogeneous, tangled magnetic field $B$. The energy distribution of the relativistic particles is assumed to follow a broken power law with indices $n_1$, $n_2$, minimum and maximum Lorentz factors $ \gamma_{\rm min}$ and $ \gamma_{\rm max}$, a break at a Lorentz factor $\gamma_{\rm br}$ and a normalisation $K_1$ at a Lorentz factor of $\gamma=$1.

When applying the usual light travel time arguments, an upper limit can be derived on the ratio of the size of the emission region (blob) $R$ and its relativistic bulk Doppler factor $\delta$ from the observed variabililty time scale $\delta t_{obs}$:
\begin{equation}
R \, \delta^{-1} < \frac{c}{1 + z} \, \Delta t_{\rm obs} \, .
\label{eq:dt}
\end{equation}

The flux variability time scale in the X-ray data of about one day constrains this ratio to $R\delta^{-1} \le$ 2.2$\times$10$^{15}$ cm for the lower redshift limit of z$=$0.176 (cf. Section~\ref{sec:intro}) and to $R\delta^{-1} \le$ 1.6$\times$10$^{15}$ cm for the upper limit of z$=$0.59, derived in Section~\ref{subsec:zlimit}. These values are not unusual for BL Lac objects.

The synchrotron and SSC peaks are not directly detected, but the peak frequencies can be constrained from the available data. The high flux level and hard spectrum in the optical and UV band, compared to the relatively low flux in the X-ray band, exclude a synchrotron peak position that is too far above the UV data and imply a peak frequency between 10$^{15}$ and 10$^{16}$ Hz in the observer's frame. 

In the Thomson limit, the Lorentz factor at the break in the electron spectrum ($\gamma_{\rm br}$) is defined by the ratio of the observed synchrotron and SSC peak frequencies, $\nu_s$ and $\nu_c$, respectively \citep{tav1998}:
\begin{equation}
\gamma_{\rm br} = \left( \frac{3 \nu_c }{4 \nu_s }  \right)^{1/2} \, .
\end{equation}
This implies a value of $\gamma_{\rm br}$ between $10^{4}$ and $10^{5}$, when assuming an SSC peak position between 10$^{24}$ and 10$^{26}$ Hz.

A minimum Doppler factor is given by the opacity condition according to \citet{tav1998} and \citet{don1995}:
\begin{equation}
\delta > \left[ \frac{\sigma_T}{5 h c^2} d_L^2 (1+z)^{2 \beta} \frac{F(\nu_0)}{\Delta t_{\rm obs}}  \right]^{1/(4+2\beta)} \, .
\end{equation}
Here  d$_L$ is the luminosity distance of the source, F$(\nu_0)$ the flux at the frequency of the target photons, $\beta$ the spectral index of the target photons and $\sigma_T$ the Thomson cross-section. For the minimum redshift of z$=$0.176, this leads to a minimum Doppler factor of $\delta \gtrsim$14. Assuming a redshift of z$=$0.3, shown 
in the following to be still acceptable for a one-zone SSC description, this value would increase to $\delta \gtrsim$18.

In the first two scenarios to be investigated, the redshift is fixed at the lower limit of z$=$0.176 and at z$=$0.20, respectively. Good representations of the SED can be found for the parameters listed in Table~\ref{tab:ssc} (first two columns); the modelled SED corresponding to z$=$0.20 (black solid line) is shown in Fig.~\ref{fig:sed1} for absorption on the EBL with the model by \citet{fra2008} and with the model by \citet{kne2010}. 
Emission at the lowest frequencies in the radio band is ascribed to the extended radio lobes (cf. Section~\ref{sec:intro}) and is not modelled here. 

To arrive at these solutions, a low magnetic field of B$=$0.02 G and relatively high bulk Doppler factors of $\delta$$=$35 and $\delta$$=$37.5 are assumed. The low magnetic field is necessary to account for the low synchrotron peak frequency. Increasing B, while keeping all other parameters fixed, shifts the synchrotron peak to higher frequencies and prevents a good representation of the X-ray slope. This can be compensated by lowering the Doppler factor accordingly, but then the overall flux level in the model decreases. To boost the flux level, the electron density K has to be increased, leading to a dominance of the SSC peak that cannot be compensated with an increase of the radius of the emission region, due to the constraint from Equation~\ref{eq:dt}.
Satisfactory solutions with lower Doppler factors can be found down to $\delta\sim$30. For smaller $\delta$, it becomes increasingly difficult to reproduce the similar flux level in the synchrotron and SSC bumps, while staying in agreement with the other constraints from the data. 

As is often seen in SSC scenarios for BL Lac objects, the energy density in electrons and positrons and the magnetic energy density are clearly out of equipartition, with  $u_e / u_B \approx 70$.

The minimum Lorentz factor of the electron distribution $\gamma_{\rm min}$ has to be of the order of a few 100 due to the upper flux limit that is provided by the radio data, although it has to be pointed out that these data are not simultaneous. The break energy of the electron distribution depends on the synchrotron and SSC peak frequencies as discussed above. 

An interpretation of the SED has also been attempted for higher redshifts, up to z$=$0.4 (cf. Table~\ref{tab:ssc} and Fig.~\ref{fig:sed1}). Only the bulk Doppler factor $\delta$
has been increased to compensate for the decreasing flux and spectral shift with higher redshift. With increasing z, absorption on the EBL becomes more severe and the model has a tendency to cut off below the H.E.S.S. spectrum. Trying to correct for this effect with an increasing break energy in the electron spectrum does not improve the result sufficiently and is limited by the position of the synchrotron peak. The model does not provide a satisfactory representation of the H.E.S.S. data for redshifts larger than about 0.35 for the EBL model by \citet{fra2008} and for redshifts larger than about 0.4 for the minimum EBL model by \citet{kne2010}, as can be seen from Fig~\ref{fig:sed1}. In addition, the Doppler factor required to compensate for the effect of increasing redshift, becomes very large for the solutions at high z. 

It can also be seen that, in the one-zone SSC scenario, the SSC peak frequency has to be located between 10$^{24}$ and 10$^{25}$ Hz.

\begin{table}[h!] 
  \centering
  \begin{tabular}{ p{0.2 \columnwidth} | l l l l l l}
    \hline
     z & 0.176 & 0.20 & 0.25 & 0.30 & 0.35 & 0.40 \\
     \hline
    $\delta$ & 35 & 37.5 & 42 & 47 & 51 & 54.5  \\ \hline
     $B$ [{\rm G}]  & \multicolumn{6}{c}{ 0.02 } \\
     $R$ [{\rm cm}]  & \multicolumn{6}{c}{ $6.5 \times 10^{16}$  } \\
     $K$ [{\rm cm}$^{-3}$] & \multicolumn{6}{c}{ 445 } \\
     $n_1$ & \multicolumn{6}{c}{ 2.0 } \\
     $n_2$ & \multicolumn{6}{c}{ 3.7 } \\
     $\gamma_\mathrm{min}$  & \multicolumn{6}{c}{ 400 } \\
     $\gamma_\mathrm{br}$  & \multicolumn{6}{c}{ $2.5 \times 10^4$ } \\
     $\gamma_\mathrm{max}$  & \multicolumn{6}{c}{ $9.0 \times 10^5$ } \\
     $u_e$ [{\rm erg cm$^{-3}$}] & \multicolumn{6}{c}{$1.6 \times 10^{-3}$} \\
    $u_B$ [{\rm erg cm$^{-3}$}] & \multicolumn{6}{c}{$2.3 \times 10^{-5}$} \\
    \hline
  \end{tabular}
  \caption{Parameters used for stationary SSC modelling of the SED of \pks for different assumptions on the redshift of the source. 
  A description of the parameters is given in the text.}
  \label{tab:ssc}
\end{table}


\section{Summary}
\label{sec:conclusions}   
   
The distant HBL \pks has been detected in the VHE band for the first time, with a significance of 15.2$\sigma$ for 13.5 hours of live time. Emission at these energies follows a steep spectrum with a photon index of 3.89$\pm$0.37 and shows marginal evidence of variability on a monthly scale. During the H.E.S.S. observations, rapid flaring was detected in the X-ray range, accompanied by a flux variation in the UV range, at a time scale of about one day. This variation is not reflected in the {\it Fermi}-LAT and VHE energy bands, but might well be hidden in the statistical uncertainties and by gaps in the coverage. This puts a limit on the size of the emission region, which is comparable to observations from other HBL.

A comparison of the {\it Fermi}-LAT spectrum with the H.E.S.S. spectrum leads to a conservative upper limit on the redshift of the source (z$<$0.59 at 95\% CL for the lower limit EBL model by \citet{gil2009}), taking into account statistical and systematic uncertainties. This is a firm upper limit, under the assumptions that the intrinsic flux does not exhibit any upturns above the {\it Fermi}-LAT band and that current models provide a correct estimate of the minimum density of the EBL. 

Assuming a simple SSC scenario to explain the SED of \pks from the optical to the VHE range, the physical parameters of the emission region and underlying particle
distribution were investigated for different redshifts. Good solutions at low redshifts require a low magnetic field and large size of the emission region, close to the upper limit from variability constraints. 
With a one-zone, homogeneous SSC model, a satisfactory description of the SED from the optical to the VHE range does not seem feasible for redshifts above about 0.4.

\begin{acknowledgements}
The support of the Namibian authorities and of the University of Namibia
in facilitating the construction and operation of H.E.S.S. is gratefully
acknowledged, as is the support by the German Ministry for Education and
Research (BMBF), the Max Planck Society, the German Research Foundation (DFG), 
the French Ministry for Research,
the CNRS-IN2P3 and the Astroparticle Interdisciplinary Programme of the
CNRS, the U.K. Science and Technology Facilities Council (STFC),
the IPNP of the Charles University, the Czech Science Foundation, the Polish 
Ministry of Science and  Higher Education, the South African Department of
Science and Technology and National Research Foundation, and by the
University of Namibia. We appreciate the excellent work of the technical
support staff in Berlin, Durham, Hamburg, Heidelberg, Palaiseau, Paris,
Saclay, and in Namibia in the construction and operation of the
equipment.

This research made use of the NASA/IPAC Extragalactic Database (NED)  and of the SIMBAD Astronomical Database. The authors thank the {\it RXTE} team for their prompt response to our ToO request and the professional interactions that followed. The authors acknowledge the use of the publicly available {\it Swift} data, as well as the public HEASARC software packages. 

The authors wish to thank H.~Landt and M.~V\'eron-Cetty for very helpful discussions on the available redshift estimates of the source.
\end{acknowledgements}

\clearpage

\appendix
\section{Treatment of the uncertainties in the {\it Fermi}-LAT flux for a derivation of the redshift upper limit}
\label{app:stats}

Apart from the approximative treatment of the systematic error in the H.E.S.S. flux, Equation~\ref{eq:tau} provides an exact expression for the 95$\%$ C.L. upper limit on the maximum optical depth $\tau_{max}$, if the intrinsic spectrum of the source is known precisely. However, if the statistical uncertainty on the {\it Fermi}-LAT spectrum is to be taken into account, an evaluation of the confidence interval is complicated by the fact that $\tau_{\rm max}$ is a function of the ratio of two Gaussian variables, the observed fluxes in the HE and VHE range, and is thus itself not a Gaussian variable. 
Moreover, the ratio of two Gaussian variables does not possess moments to any order greater or equal to one.
Therefore, confidence intervals must rely on percentiles.

As will be shown in the following, the confidence interval $[ D_{-n}, D_{+n}]$ at the $n$-$\sigma$ confidence level for the ratio of two independent Gaussian variables with mean $\alpha$, $\beta$ and standard deviation 
$\sigma_{\alpha}$, $\sigma_{\beta}$ is given by the expression:

\begin{equation}
D_{\pm n} = \frac{  \alpha \pm \frac{\beta^*}{(\alpha^{*2} + \beta^{*2} - n^2)^{\frac{1}{2}} } \; n \sigma_{\alpha}  }{ \beta \mp  \frac{\alpha^*}{(\alpha^{*2} + \beta^{*2} - n^2)^{\frac{1}{2}} } \; n \sigma_{\beta}  }\,.
\label{eq.A1}
\end{equation}

Here, $\alpha^* \equiv \alpha /  \sigma_{\alpha}$, $\beta^* \equiv \beta /  \sigma_{\beta}$ and $n$ is the $n$-$\sigma$ parameter defining the confidence level $\gamma$ (see Table~\ref{tab.A1} for a few examples of $\gamma$ as a function of $n$).
This can be seen as follows:
suppose one is interested in the ratio $D=y/x$ of two independent Gaussian random variables $x$ and $y$ : $x\sim{\cal N}(\beta,\sigma_\beta^2)$, $y\sim{\cal N}(\alpha,\sigma_\alpha^2)$.
The joint probability distribution function of the couple $(x,y)$ is
\begin{equation}
f(x,y)=
\frac{1}{2\pi}
\exp \{ -\frac{1}{2}\Bigl[\Bigl(\frac{x-\beta}{\sigma_\beta}\Bigr)^2+\Bigl(\frac{y-\alpha}{\sigma_\alpha}\Bigr)^2\Bigr]\, \} \, .
\end{equation}
Next, the change of variables $x=\rho\sigma_\beta\cos\phi$ and $y=\rho\sigma_\alpha\sin\phi$ is performed.
Knowing $\phi$, one knows the ratio $D$, therefore one seeks the probability distribution function $f_\phi$ of $\phi$, which is
\begin{equation}
f_\phi(\phi)=
\frac{1}{2\pi}e^{-\frac{1}{2}R^2}\int_0^\infty\!\rho e^{-\frac{1}{2}\rho^2}e^{R\rho\cos(\phi-\phi_0)}\,d\rho\,.
\label{eq.A3}
\end{equation}
with
\begin{equation}
R=\Bigl[\frac{\alpha^2}{\sigma_\alpha^2}+\frac{\beta^2}{\sigma_\beta^2}\Bigr]^{\frac{1}{2}}\,,
\quad
\tan\phi_0=\frac{\alpha}{\sigma_\alpha}\frac{\sigma_\beta}{\beta}\,.
\end{equation}
The integral in~\eqref{eq.A3} can be evaluated easily. It yields
\begin{equation}
\begin{split}
&f_\phi(\phi)=
\frac{1}{2\pi}\bigl[e^{-\frac{1}{2}R^2}+\\
&+\sqrt{2\pi}R\cos(\phi-\phi_0)e^{-\frac{1}{2}R^2\sin^2(\phi-\phi_0)}\Phi(R\cos(\phi-\phi_0))\bigr]\,,
\end{split}
\label{eq.A4}
\end{equation}
where $\Phi$ is the cumulative distribution function of a reduced Gaussian variable, that is
\begin{equation}
\Phi(t)\equiv\int_{-\infty}^t\frac{1}{\sqrt{2\pi}}e^{-\frac{1}{2}u^2}\,du\,.
\end{equation}
From now on, two approximations are introduced.
First, the signal-to-noise ratio $R$ is supposed to be sufficiently high, such that $e^{-\frac{1}{2}R^2}\approx 0$.
Second, the variations of $\phi$ about $\phi_0$ are supposed not to be too extreme, such that $\Phi(R\cos(\phi-\phi_0))\approx1$.
This last condition is certainly fulfilled if all quantities of interest stay positive.
Under this condition, the trigonometric functions are one-to-one relations of their arguments.
Equation~\eqref{eq.A4} reduces to
\begin{equation}
f_\phi(\phi)=
\frac{1}{\sqrt{2\pi}}R\cos(\phi-\phi_0)\exp\bigl[-\tfrac{1}{2}R^2\sin^2(\phi-\phi_0)\bigr]\,.
\end{equation}
If one sets $t=R\sin(\phi-\phi_0)$, the probability distribution function (p.d.f.) of $t$ equals
\begin{equation}
f_t(t)=\frac{1}{\sqrt{2\pi}}\exp\bigl(-\tfrac{1}{2}t^2\bigr)\,,
\end{equation}
which is the p.d.f. of a reduced Gaussian variable.
Since there are one to one relations between $t$, $\phi$ and the ratio $D$, the percentiles of $D$ are uniquely determined by the percentiles of $t$.
Next, one defines $t_n$, $t_{-n}\equiv -t_n$ and the confidence level $\gamma$ by
\begin{equation}
\int_{t_{-n}}^{t_n}\!\frac{1}{\sqrt{2\pi}}\exp\bigl(-\tfrac{1}{2}t^2\bigr)\,dt=\Phi(n)-\Phi(-n)=\gamma(n)\,.
\end{equation}
That is
\begin{equation}
t_{\pm n}=\pm n\,.
\end{equation}
From this one derives successively
\begin{gather}
R\sin(\phi_{\pm n}-\phi_0)=\pm n\,,\\
\phi_{\pm n}-\phi_0=\arcsin\frac{\pm n}{R}\,,\\
\phi_{\pm n}=\phi_0\pm\arcsin\frac{n}{R}\,,\\
\arctan\Bigl(D_{\pm n}\frac{\sigma_\beta}{\sigma_\alpha}\Bigr)
=
\arctan\Bigl(\frac{\alpha}{\beta}\frac{\sigma_\beta}{\sigma_\alpha}\Bigr)
\pm
\arcsin\frac{n}{R}\,,\\
D_{\pm n}\frac{\sigma_\beta}{\sigma_\alpha}
=
\tan(\arctan\Bigl(\frac{\alpha}{\beta}\frac{\sigma_\beta}{\sigma_\alpha}\Bigr)
\pm
\arcsin\frac{n}{R}
)\,.
\end{gather}
Now, with the use of the trigonometric relations: $\tan(a\pm b)=(\tan a\pm \tan b)/(1\mp\tan a\tan b)$ and $\tan\arcsin x=x/\sqrt{1-x^2}$, one gets
\begin{equation}
D_{\pm n}=\frac
{\frac{\alpha}{\beta}\pm\frac{\sigma_\alpha}{\sigma_\beta}\frac{n}{(R^2-n^2)^{\frac{1}{2}}}}
{1\mp\frac{\alpha}{\beta}\frac{n}{(R^2-n^2)^{\frac{1}{2}}}}\,.
\end{equation}
Introducing $\alpha^* \equiv \alpha /  \sigma_{\alpha}$ and $\beta^* \equiv \beta /  \sigma_{\beta}$, such that $R^2=\alpha^{*2}+\beta^{*2}$, one finds Equation~\eqref{eq.A1}.

\begin{table}
\centering
\begin{tabular}{c|ccccc}
$n$ & 1 & 1.64 & 2 & 2.58 & 3 \\
\hline
$\gamma$ & 68.3\% & 90.0\% & 95.4\% & 99.0\% & 99.7\%
\end{tabular}
\caption{Confidence level $\gamma$ as a function of the $n$-$\sigma$ parameter for typical values of $n$ and $\gamma$.}
\label{tab.A1}
\end{table}

 \begin{figure}
     \centering
      \includegraphics[width=\columnwidth]{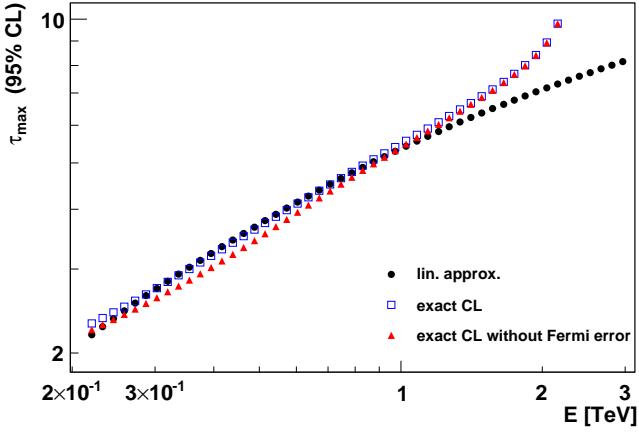}
  \caption{ Distribution of $\tau_{max}$ vs. energy for three different treatments of the statistical errors (systematic errors are neglected).
  red triangles: only statistical errors in the H.E.S.S. spectrum are taken into account;
  black circles: statistical errors in the {\it Fermi}-LAT and H.E.S.S. fluxes are included in the linear approximation;
  blue open squares: statistical errors from {\it Fermi}-LAT and H.E.S.S. are included using the exact analytic solution derived here.}
  \label{fig:taumax_nosyserr}
\end{figure}

The effect of including the uncertainties on the {\it Fermi}-LAT spectrum on the resulting distribution of $\tau_{\rm max}(E)$ can be seen from Fig.~\ref{fig:taumax_nosyserr}, where the exact result derived here, i.e. D$_{+n}$ with n$=$1.64, is compared to the 
distribution arrived at when neglecting the statistical uncertainty in the {\it Fermi}-LAT spectrum, and with an approach where H.E.S.S. and {\it Fermi}-LAT errors are
treated in the linear approximation, i.e. by simple error propagation. Systematic uncertainties have been neglected here. It should be noted that in the energy range of 
interest for the derivation of an upper redshift limit, around 1 TeV, all three approaches provide very similar results.

 \begin{figure}
     \centering
      \includegraphics[width=\columnwidth]{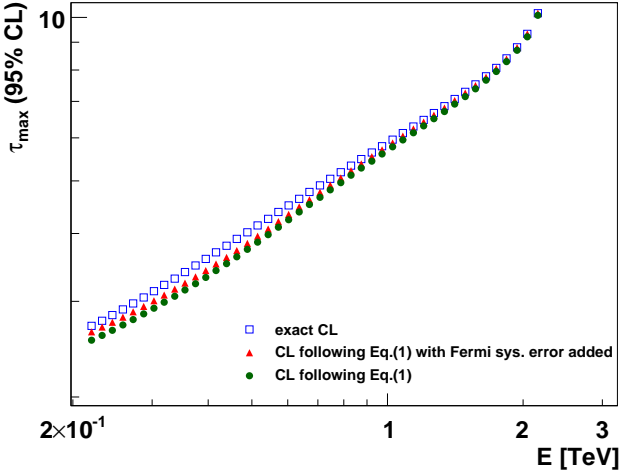}
  \caption{ Distribution of $\tau_{\rm max}$ vs. energy including systematic errors as flux biases. 
  green circles: identical to the curve in Fig.~\ref{fig:zlimit1}, following Eq.~\ref{eq:tau}; 
  red triangles: same as above, but with a systematic error of 10$\%$ added to the {\it Fermi}-LAT spectrum (cf. Sec.~\ref{subsec:mwl_fermi});
  blue open circles: statistical errors from {\it Fermi}-LAT and H.E.S.S. are included using the exact solution derived here, systematic errors have been added to both fluxes as well.}
  \label{fig:taumax_syserr}
\end{figure}

For the curves shown in Fig.~\ref{fig:taumax_syserr}, an estimate of the systematic uncertainty in the H.E.S.S. and {\it Fermi}-LAT fluxes has been added as a bias, 
as described in Sec.~\ref{subsec:zlimit} for H.E.S.S. For {\it Fermi}-LAT, a systematic bias of 10$\%$ has been assumed. In the energy range of interest, the exact solution is in good agreement with the curve used to derive the upper limit on the redshift of the source (cf. Fig.~\ref{fig:zlimit1}). The systematic bias added to the {\it Fermi}-LAT flux and the inclusion of the
statistical uncertainties in the {\it Fermi}-LAT flux have a negligible effect.


\begin{thebibliography}{99}

\bibitem[Abdo et al.(2009)]{abd2009} Abdo,~A.~A., Ackermann,~M., Ajello,~M. et al.  (Fermi-LAT Collaboration), 2009, ApJ, 700, 597 
\bibitem[Abdo et al.(2010)]{abd2010} Abdo,~A.~A., Ackermann,~M., Ajello,~M. et al. (Fermi-LAT Collaboration), 2010, ApJ, 708, 1310
\bibitem[Abramowski et al.(2012)]{abr2012} Abramowski,~A., Acero,~F., Aharonian,~F. et al. (H.E.S.S. Collaboration), 2012, A\&A, 539, 149
\bibitem[Ackermann et al.(2011)]{ack2011} Ackermann,~M., Ajello,~M., Allafort,~A. et al. (Fermi-LAT Collaboration), 2011, ApJ, 743, 171
\bibitem[Aharonian et al.(2002)]{aha2002}Aharonian,~F.~A., Timokhin,~A.~N. \& Plyasheshnikov,~A.~V., 2002, A\&A, 384, 834
\bibitem[Aharonian et al.(2006)]{aha2006} Aharonian,~F.~A., Akhperjanian, A.~G., Bazer-Bachi, A.~R., et al. (H.E.S.S. collaboration), 2006, A\&A, 457, 899
\bibitem[Aharonian et al.(2012)]{aha2012} Aharonian,~F., Essey,~W., Kusenko,~A., et al., 2012, astro-ph/1206.6715 
\bibitem[Akerlof et al.(2000)]{ake2000} Akerlof, C. W., Balsano, R., Barthelmy, S., et al. 2000, ApJ, 542, 251
\bibitem[Akerlof et al.(2003)]{ake2003} Akerlof, C. W., Kehoe, R. L., McKay, T. A., et al. 2003, PASP, 115, 132
\bibitem[Aleksi\'{c} et al.(2011)] {ale2011}Aleksi\'{c},~J., Antonelli,~L., Antoranz,~P. et al., 2011, ApJ Lett., 730, L8
\bibitem[Atwood et~al. (2009)]{atw2009}Atwood,~W.~B., Abdo,~A.~A., Ackermann,~M. et al. (Fermi-LAT collaboration), 2009, \apj, 697, 1071
\bibitem[Berge et al.(2007)]{ber2007} Berge,~D., Funk,~S. \& Hinton,~J., 2007, A\&A, 466, 1219 
\bibitem[Burrows et al.(2005)]{bur2005} Burrows,~D.~N., Hill,~J.~E., Nousek,~J.~A., et al. 2005, Space Sci. Rev., 120, 165
\bibitem[Campana et al.(2006)]{cam2006} Campana,~S., Beardmore,~A.~P., Cusumano,~G., et al., 2006, Swift XRT CALDB Release Note 09: Response Matrices and Ancillary Response Files (Washington, DC: NASA), http://heasarc.gsfc.nasa.gov/docs/heasarc/caldb/swift/docs/xrt/SWIFT-XRT-CALDB-09.pdfÊ
\bibitem[Craig \& Fruscione(1997)]{cra1997} Craig,~N. \& Fruscione,~A., 1997, Astron. J. 114, 1356 
\bibitem[De Angelis, Galanti \& Roncadelli(2011)]{dea2011} De~Angelis,~A., Galanti,~G. \& Roncadelli,~M., 2011, Phys. Rev. D, 84, 105030
\bibitem[Dondi \& Ghisellini(1995)]{don1995} Dondi,~L. \& Ghisellini,~G., 1995, MNRAS, 273, 583
\bibitem[Donea \& Protheroe(2003)]{don2003} Donea,~A.-C. \& Protheroe,~R.~J., 2003, Astropart. Phys., 18, 377  
\bibitem[Essey et al.(2011)]{ess2011} Essey,~W., Kalashev,~O., Kusenko,~A. et al., 2011, ApJ, 731, 51
\bibitem[Feldman \& Cousins(1998)]{fel1998} Feldman,~G.~J., \& Cousins~R.~D., 1998, Phys. Rev. D, 57, 3873
\bibitem[Franceschini et al.(2008)]{fra2008} Franceschini,~A., Rodighiero,~G., \& Vaccari,~M., 2008, A\&A, 487, 837 
\bibitem[Fumagalli et al. (2012)]{fum2012} Fumagalli,~M., Furniss,~A., O'Meara,~J. et al., 2012, A\&A, 545, 68
\bibitem[Georganopoulos, Finke \& Reyes(2010)]{geo2010} Georganopoulos,~M., Finke,~J.~D. \& Reyes,~L.~C.,  2010, ApJ, 714, 157
\bibitem[Gilmore et al.(2009)]{gil2009} Gilmore,~R.~C., Madau,~P., Primack,~J.~R.  et al., 2009, MNRAS, 399, 1694
\bibitem[Gregory et al.(1994)]{gre1994} Gregory,~P.~C., Vavasour,~J.~D. \& Scott,~W.~K., 1994, ApJS, 90, 173 
\bibitem[Haakonsen \& Rutledge(2009)]{haa2009} Haakonsen,~C.~B. and Rutledge,~R.~E., 2009, ApJS, 184, 138 
\bibitem[Hauser et al.(2004)]{hau2004} Hauser,~M., M{\"o}llenhoff,~C. \& P{\"u}hlhofer,~G., 2004 Astronomische Nachrichten 325, 659
\bibitem[Healey et al.(2007)]{hea2007} Healey, ~S.~E., Romani,~R.~W., Taylor,~G.~B., et al., 2007, ApJS, 171, 61
\bibitem[Jahoda et al.(1996)]{jah1996} Jahoda, K. et al., 1996, in ``EUV, X-ray and Gamma-ray Instrumentation for Astronomy VII'', SPIE Proc., 2808, 59
\bibitem[Kalberla et al.(2005)]{kal2005} Kalberla, P.~M.~W., Burton, W.~B., Hartmann, D., et al.\ 2005, \aap, 440, 775
\bibitem[Katarzynski, Sol \& Kus (2001)]{Kat2001}Katarzy{\'n}ski,~K., Sol,~H. \& Kus, A., 2001, A\&A, 367, 809
\bibitem[Kneiske \& Dole (2010)]{kne2010} Kneiske,~T.~M. \& Dole,~H., 2010, A\&A, 515, 19
\bibitem[Lampton et al.(1997)]{lam1997} Lampton,~M., Lieu,~R., Schmitt,~J.~H.~M.~M. et al., 1997, ApJS, 108, 545 
\bibitem[Landt(2012)]{lan2012}Landt,~H., 2012, MNRAS Lett., 423, 84
\bibitem[Landt \& Bignall(2008)]{lan2008} Landt,~H. \& Bignall,~H.~E., 2008, MNRAS, 391, 967 
\bibitem[Large et al.(1981)]{lar1981} Large,~M.-I., Mills,~B.~Y., Little,~A.~G. et al., 1981, MNRAS, 194, 693 
\bibitem[Lefa, Aharonian \& Rieger (2011)]{lef2011} Lefa,~E., Aharonian,~F.~A., \& Rieger,~F.~M., 2011, ApJ, 743, L19
\bibitem[Mattox et~al. (1996)]{matt1996}Mattox, J.~R., et~al., 1996, \apj, 461, 396
\bibitem[Mauch et~al.(2003)]{mau2003}Mauch,~T., Murphy,~T., Buttery,~H.~J. et al., 2003, MNRAS, 342, 1117
\bibitem[McConnel et~al.(2012)]{mcc2012}McConnell,~D., Sadler~,E.~M., Murphy~T., et al., 2012, MNRAS, 422, 1527
\bibitem[Murase et al.(2012)]{mur2012} Murase,~K., Dermer,~C.~D., Takami,~H. et al., 2012, ApJ, 749, 63
\bibitem[Murphy  et~al.(2010)]{mur2010}Murphy,~T., Sadler,~E.~M., Ekers~R.~D. et al., 2010, MNRAS, 402, 2403
\bibitem[de Naurois \& Rolland(2009)]{den2009} de Naurois,~M. \& Rolland,~L., 2009, Astropart. Phys., 32, 231 
\bibitem[Neronov, Semikoz \& Taylor]{ner2011} Neronov,~A., Semikoz,~D. \& Taylor,~A.~M., 2011, A\&A, 541, 31
\bibitem[Padovani \& Giommi(1995)]{pad1995}Padovani,~P. \& Giommi,~P., 1995, MNRAS, 277, 1477
\bibitem[Perlman et al.(1998)]{per1998} Perlman,~E.~S. et al., 1998, ApJ, 115, 1253 
\bibitem[Piranomonte et al.(2007)]{pir2007}Piranomonte,~S., Perri,~M., Giommi,~P. et al., 2007, A\&A, 470, 787 
\bibitem[Piron et al.(2001)]{pir2001} Piron,~F. et al. 2001, A\&A, 374, 895
\bibitem[Pita et al.(2012)]{pit2012} Pita,~S., Goldoni,~P., Boisson,~C. et al. 2012, proceedings of the Gamma 2012 conference, astro-ph/1208.1785
\bibitem[Poole et al.(2008)]{poo2008} Poole,~T.~S., Breeveld,~A.~A., Page,~M.~J. et al., 2008, MNRAS, 383, 627
\bibitem[Poutanen \& Stern (2012)]{pou2012} Poutanen,~J. \& Stern,~B.~E., 2012, PoS(AGN 2011)015, astro-ph/1109.0946
\bibitem[Prandini et al.(2010)]{pra2010} Prandini,~E., Bonnoli,~G., Maraschi,~L. et al., 2010, MNRAS, 405L, 76
\bibitem[Prandini, Bonnoli \& Tavecchio (2011)]{pra2011} Prandini,~E., Bonnoli,~G. \& Tavecchio,~F., 2012, A\&A, 543, 111
\bibitem[Raue et al.(2009)]{rau2009} Raue,~M. et al. (for the H.E.S.S. Collaboration), ATel 2350 (2009)
\bibitem[Roming et al.(2005)]{rom2005} Roming,~P.~W.~A., Kennedy,~T.~E., Mason,~K.~O. et al., 2005, Space Sci. Rev., 120, 95
\bibitem[Roming et al.(2009)]{rom2009} Roming,~P.~W.~A., Koch, T.~S., Oates, S.~R. et al., 2009, ApJ, 690, 163
\bibitem[Schlegel et al.(1998)]{sch1998} Schlegel,~D.~J. et al., 1998, ApJ, 500, 525 
\bibitem[Senturk et al.(2011)]{sen2011} Senturk,~G.~D., Errando,~M., Boettcher,~M. et al., 2011, proceedings of the Fermi symposium, eConf C110509, astro-ph/1111.0378
\bibitem[Tavecchio et al.(1998)]{tav1998} Tavecchio,~F., Maraschi,~L. \& Ghisellini,~G., 1998, ApJ, 509, 608
\bibitem[Taylor et al.(2011)]{tay2011} Taylor,~A., Vovk,~I. \& Neronov,~A., 2011, A\&A, 529, 144
\bibitem[Urry et al.(2000)]{urr2000} Urry,~C.~M., Scarpa,~R., O'Dowd,~M. et al., 2000, ApJ, 532, 816 
\bibitem[Vaughan et al.(2003)]{vau2003}Vaughan,~S., Edelson,~R., Warwick,~R.~S. et al., 2003, MNRAS, 345, 1271
\bibitem[Veron-Cetty \& Veron (2010)]{ver2010} Veron-Cetty,~M.-P. and Veron,~P., 2010, A\&A, 518, 10
\bibitem[White et al.(1994)]{whi1994}White,~N.~E., Giommi,~P, Angelini,~L., 1994, proceedings of the HEAD meeting 1994
\bibitem[Wright \& Otrupcek (1990)]{wri1990} Wright,~A. \& Otrupcek,~R., 1990, Parkes Catalog
\bibitem[Wright et al.(1994)]{wri1994} Wright,~A., Griffith,~M.~R., Burke,~B.~F. et al., 1994, ApJS, 91, 111
\bibitem[Zacharopoulou et al.(2011)]{zac2011} Zacharopoulou,~O., Khangulyan,~D., Aharonian,~F.~A., et al., 2011, \apj, 738, 157 
\bibitem[Zech et al.(2011)]{zec2011}Zech,~A., Behera,~B, Becherini,~Y. et al. (for the H.E.S.S. Collaboration), 2011, PoS (Texas 2010) 200, astro-ph/1105.0840
\bibitem[Zhang et al.(1999)]{zha1999}Zhang,~Y.~H., Celotti,~A., Treves,~A. et al., 1999, ApJ 527, 719
\end{thebibliography}
\end{document}